\documentclass[journal]{IEEEtran}
\usepackage{cite}
\usepackage{amsmath,amssymb,amsfonts}
\usepackage{algorithmic}
\usepackage{graphicx}
\usepackage[caption=false]{subfig}
\usepackage{textcomp}
\usepackage{multirow}
\usepackage{color, soul}
\usepackage{enumitem}
\usepackage{import}
\usepackage[ruled,vlined]{algorithm2e}
\usepackage{setspace}
\usepackage{hyperref}
\usepackage{tabularx}
\usepackage{comment}
\usepackage{amssymb,graphicx,stackengine,xcolor}
\usepackage{chngpage}
\usepackage{balance}
\usepackage[center]{caption}

\usepackage{tikz}

\newcommand{\pie}[1]{%
\begin{tikzpicture}
 \draw (0,0) circle (1ex);\fill (1ex,0) arc (0:#1:1ex) -- (0,0) -- cycle;
\end{tikzpicture}%
}
\ifCLASSINFOpdf
\else
\fi

\begin{document}

\title{Consumer, Commercial and Industrial IoT (In)Security: Attack Taxonomy and Case Studies} 

\author{Christos~Xenofontos*,~\IEEEmembership{Student Member,~IEEE,}
        Ioannis~Zografopoulos*,~\IEEEmembership{Graduate Student Member,~IEEE,}~Charalambos~Konstantinou*,~\IEEEmembership{Senior~Member,~IEEE}, Alireza Jolfaei,~\IEEEmembership{Senior~Member,~IEEE}, Muhammad Khurram Khan,~\IEEEmembership{Senior~Member,~IEEE}, and Kim-Kwang Raymond Choo,~\IEEEmembership{Senior~Member,~IEEE}

\thanks{C. Xenofontos is with the Department of Applied Mathematics and Computer Science, Richard Petersens Plads, Technical University of Denmark (DTU), Kongens Lyngby 2800, Denmark (e-mail: cxenof03@ieee.org). \protect 

I. Zografopoulos and C. Konstantinou are with the Division of Computer, Electrical and Mathematical Sciences and Engineering, King Abdullah University of Science and Technology (KAUST), Thuwal 23955, Saudi Arabia (e-mail: jzographopoulos@gmail.com, ckonstantinou@ieee.org). \protect 

A. Jolfaei is with the Department of Computing, Macquarie University, Sydney NSW 2113, Australia (e-mail: alireza.jolfaei@mq.edu.au).
\protect 

M. K. Khan is with the Center of Excellence in Information Assurance (CoEIA), King Saud University,
Saudi Arabia (e-mail: mkhurram@ksu.edu.sa). \protect 

K.-K. R. Choo is with the Department of Information Systems and Cyber Security, University of Texas at San Antonio (UTSA), 1 UTSA Circle, San Antonio, TX 78249-0631, USA. He also has courtesy appointments at UTSA’s Department of Electrical and Computer Engineering and Department of Computer Science, and UniSASTEM, University of South Australia, Adelaide, SA 5095, Australia. (e-mail: raymond.choo@fulbrightmail.org).
\protect \\
*Work partly performed while these authors were with the Center of Advanced Power Systems (CAPS), Florida State University (FSU). \protect \\
Copyright (c) 20xx IEEE. Personal use of this material is permitted. However, permission to use this material for any other purposes must be obtained from the IEEE by sending a request to pubs-permissions@ieee.org.}
}

\markboth{}%
{Xenofontos \MakeLowercase{\textit{et al.}}: }

\IEEEaftertitletext{\vspace{-1.5\baselineskip}}

\maketitle
\begin{abstract}
Internet of Things (IoT) devices are becoming ubiquitous in our lives, with applications spanning from the \emph{consumer} domain to \emph{commercial} and \emph{industrial} systems. The steep growth and vast adoption of IoT devices reinforce the importance of sound and robust cybersecurity practices during the device development life-cycles. IoT-related vulnerabilities, if successfully exploited can affect, not only the device itself, but also the application field in which the IoT device operates. Evidently, identifying and addressing every single vulnerability is an arduous, if not impossible, task. Attack taxonomies can  assist in classifying attacks and their corresponding vulnerabilities. Security countermeasures and best practices can then be leveraged to mitigate threats and vulnerabilities before they emerge into catastrophic attacks and ensure overall secure IoT operation.  Therefore, in this paper, we provide an attack taxonomy which takes into consideration the different layers of IoT stack, i.e., device, infrastructure, communication, and service, and each layer's designated characteristics which can be exploited by adversaries. Furthermore, using nine real-world cybersecurity incidents, that had targeted IoT devices deployed in the consumer, commercial, and industrial sectors, we describe the IoT-related vulnerabilities, exploitation procedures, attacks, impacts, and potential mitigation mechanisms and protection strategies. These (and many other) incidents highlight the underlying security concerns of IoT systems and demonstrate the potential attack impacts of such connected ecosystems, while the proposed taxonomy provides a systematic procedure to categorize attacks based on the affected layer and corresponding impact. 
\end{abstract}


\IEEEpeerreviewmaketitle

\vspace{-0.2in}
\section{Introduction} \label{s:Intro}

The number of Internet of Things (IoT) devices keeps increasing. By the end of 2030, the number of connected devices is expected to reach 24.1 billion, compared with around 500 million devices in 2003, which corresponds to around 3.47 IoT devices per person \cite{transforma,CISCO}. Out of the 24.1 billion devices, it is estimated that 5.8 billions will be allocated only for enterprises and industrial applications \cite{Gartner}. These numbers highlight the importance of IoT as people and devices are drastically transforming the way they measure, sense, and communicate with their connected ecosystems. The extensive deployment of IoT devices, however, raises security concerns. Given the plurality of IoT architectures, incorporating a plethora of sensing and communication modules, integrating such devices results in complex, and dynamic landscape \cite{8796409}.

A 2020 study conducted by Nokia's threat intelligence labs, for example, indicated that IoT devices account for almost 30\% of the attacks encountered in mobile and wireless networks (e.g., WiFi, Bluetooth, etc.) \cite{IOT_sec}. In many circumstances, the security of IoT devices is often  constrained by their application field. For instance, it has been demonstrated that handheld, portable, and wearable devices -- relying on battery-powered operation -- often trade-off security performance to sustain longer operation time. Despite the development of energy efficient communication and control protocols for IoT to overcome such issues, vulnerabilities in such protocols can still be stealthily lurking. A prominent example affecting billions of IoT devices is the Bluetooth low energy~(BLE) communication protocol. BLE is extensively used for both wearable as well as industrial IoT (IIoT) ecosystems. It  has been found vulnerable to multiple attacks. Researchers have demonstrated that BLE can expose user data since typically the transmitted packets are exchanged in plain text. Also, BLE includes security flaws in the device authentication phase, i.e., after two paired devices are reconnected \cite{wu2020blesa, antonioli2020blurtooth}.

\begin{figure}[t]

\centerline{\includegraphics[width=\linewidth]{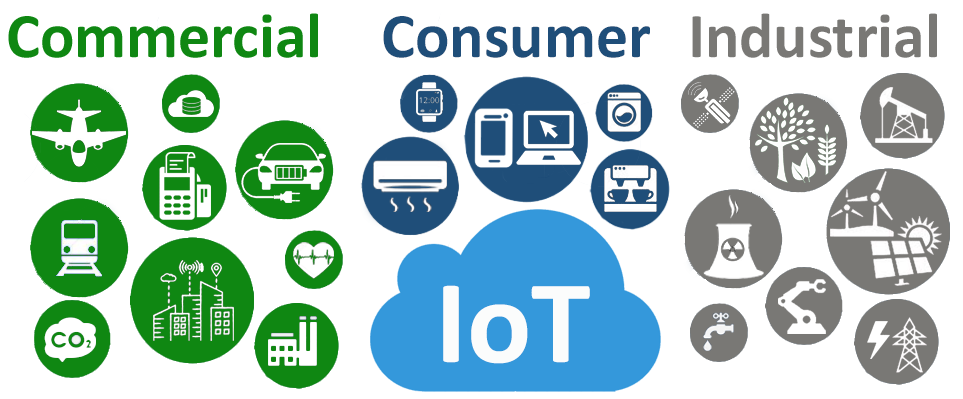}}
\caption{Overview of commercial, consumer, and industrial Internet of Things (IoT) sectors.}
\label{fig:IoT}
\end{figure} 

Attackers can leverage IoT vulnerabilities and mount their attacks in any IoT ecosystem. In Fig. \ref{fig:IoT}, we depict such IoT environments including smart watches and mobile phones, connected medical devices, and smart cars, but also smart cities, transportation systems, manufacturing industries, and critical infrastructures.  The severity of IoT attacks depends on the targeted system, the compromised asset (e.g., mobile phone, medical device, or smart grid controller), and the information stored or utilized for the asset's operation and control. Most, if not all, IoT ecosystems such as IIoT\cite{9146364}, Internet of Energy~\cite{7891908}, Internet of Medical Things~\cite{9298452}, military IoT~\cite{9317401}, etc., have underlying hidden or unknown vulnerabilities that can have diverse consequences if they are successfully exploited by malicious threat actors, such as nation states and advanced persistent threat (APT) adversaries. Although previous works have underlined the security implications of IoT devices  \cite{chen2018internet, alaba2017internet, 7804660, 9219584}, in this work, we systematically review IoT security from three major sectors, i.e., \textit{(i)} consumer, \textit{(ii)} commercial, and \textit{(iii)} industrial. The rationale for this distinction stems from the diverse operational requirements, implicit security constraints, mission criticality, and potential outcomes in the event of a compromise targeting the respective IoT sectors. In the following, we provide definitions for each one of the three IoT sectors.

\begin{itemize}[leftmargin=*, wide=0pt]

\item\emph{Consumer IoT:} targets end-user applications and includes personal devices such as smartphones, smartwatches, and other wearables. Internet-connected home devices (e.g., thermostats, cameras, smart lamps, etc.) and appliances (e.g., A/C, heaters, refrigerators, etc.) able to  collect data and be remotely monitored and controlled are also considered consumer IoT devices.

\item\emph{Commercial IoT:} refers to the resources utilized by enterprises (e.g., offices, health monitoring systems, storage facilities, etc.) as well as bigger infrastructures (e.g., smart city deployments, transportation and electric vehicle monitoring, communication and control, etc.). Commercial IoT devices are used to automate, coordinate, and respond to changes in the commercial environment, while minimizing operational cost and service latency.

\item\emph{Industrial IoT:} includes sensors, actuators, controllers, industrial assets (e.g., manufacturing robotics, power plant controllers, etc.), remote telemetry, monitoring and management systems (e.g., human-machine interfaces -- HMIs, remote terminal units -- RTUs, etc.). This interconnected mission-critical architecture enables the real-time exchange of industrial system information, providing better situational awareness, improving the control over system processes, and increasing productivity and efficiency.

\end{itemize}

IoT devices, irrespective of their application or operating environment, are responsible for monitoring, controlling and enhancing system connectivity and performance. In terms of fundamental security objectives, IoT device operation needs to ensure the confidentiality of the sensed measurements, safeguard the information integrity of stored or in-transit data, and grant access only to authorized users/parties \cite{7562568}.  Although, security investigations revolve around the confidentiality-integrity-availability (CIA) principle, the order in which security objectives are prioritized differs significantly depending on the IoT sector. For instance, the confidentiality of the data residing in consumer IoT devices is of most importance for the device users. 
On the other hand, in a smart city deployment, for instance,  the availability of the respective IoT nodes arises as the most important objective. 
In industrial automation environments including critical infrastructure, e.g., nuclear plants, the integrity of the sensed data and availability of system resources are of utmost significance since they can lead to uneconomical plant operation or even catastrophic incidents. 

Due to the interoperability,  heterogeneous architectures, and distinct security objectives of IoT devices and the sector in which they are installed, comprehensive security studies are required to address and evaluate vulnerabilities (on every layer of the IoT stack) and identify potential attack entry points. Towards contributing to this task, in this paper, we propose an attack taxonomy factoring  the unique structures and operational constraints of IoT ecosystems. Furthermore, we report nine IoT-based attack incidents, three for each of the attacked IoT sectors, i.e., consumer, commercial, and industrial. We explain the attack vectors that adversaries can potentially exploit to mount their attacks as well as the impacts on the system operation. Based on our proposed taxonomy, we demonstrate how these IoT attacks can be mapped to their respective categories and discuss potential mitigation strategies.

The rest of the paper is organized as follows. Section \ref{s:taxonomy} discusses related work and introduces our proposed IoT attack taxonomy. Section \ref{s:attackIncidents} describes real-world incidents, adversarial attack paths, and mitigation strategies for cyberattacks in consumer, commercial, and industrial IoT deployments. Section \ref{s:challenges} discusses open challenges and research directions in the field of IoT security. Finally, Section \ref{s:conclusion} concludes the paper and summarizes the lessons learned from the attack incidents and provides directions for future improvements in the field of IoT security.

\section{IoT Attack Taxonomy} \label{s:taxonomy}

\begin{figure}[t]
\centerline{\includegraphics[width=\linewidth]{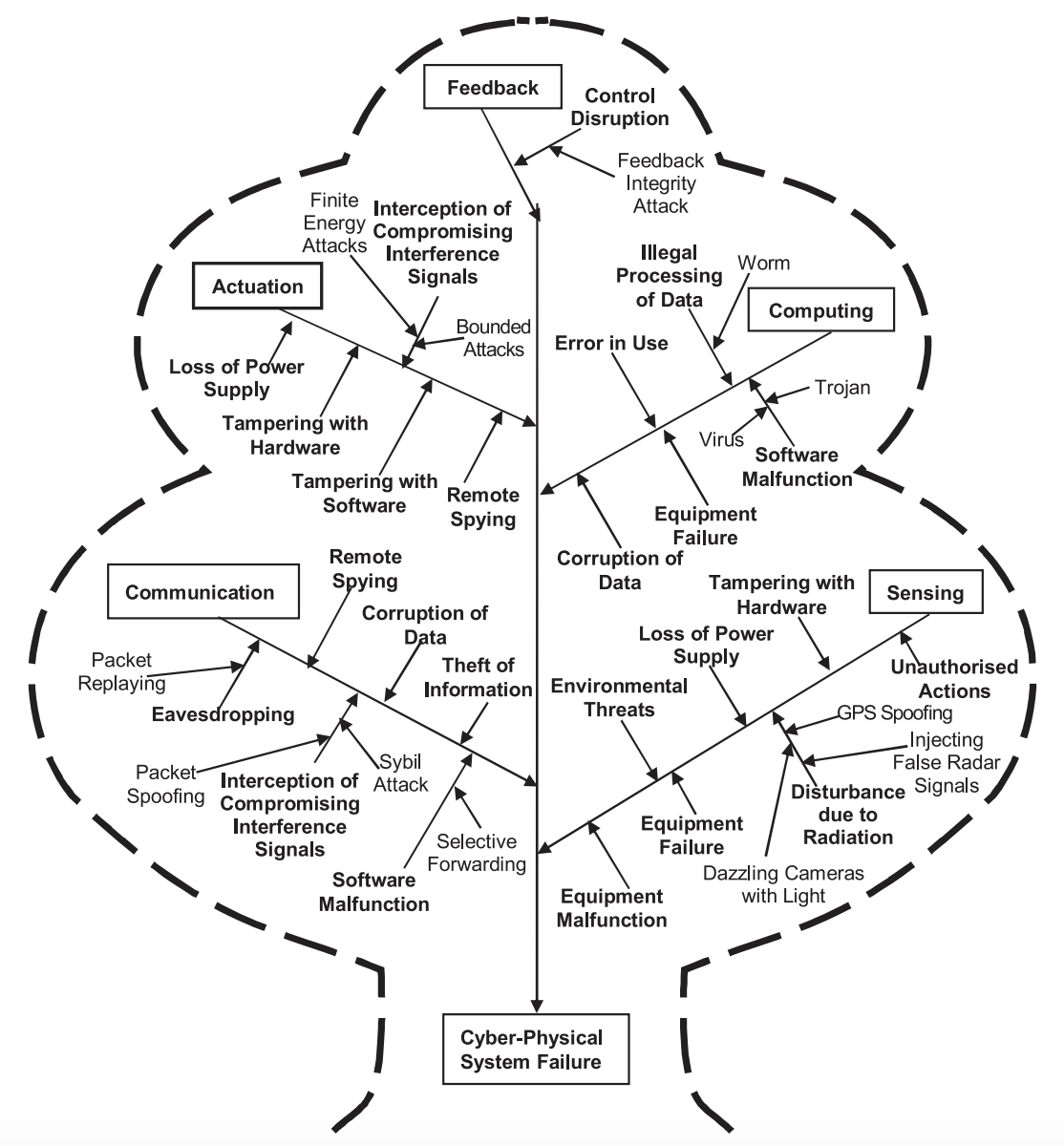}}
\caption{A tree diagram of attacks and threats on Internet of Things (IoT) and cyber-physical systems (CPS) \cite{alguliyev2018cyber}. }
\label{fig:attackleaf}
\end{figure}

{Many diverse technologies such as wide-area networks, data analytics, security platforms, and operating systems are involved in the IoT spectrum, and as a result, a plethora of the reported attacks targeting IoT devices are directed to such technologies \cite{dosal_2020}. Attack taxonomies can assist in investigating IoT attacks, focusing on the layer where an attack is materialized overcoming drawbacks related to attack classification methods focusing on specific device- or technology-induced vulnerabilities. In this part, we review existing IoT taxonomies and introduce our proposed layered classification while illustrating  how real-world IoT cyberattacks can fit into our taxonomy.  We survey prominent papers in the IoT security field published in top-tier journal and conferences from 2015-2020, emphasizing in articles that consist of different IoT taxonomies. The literature review was conducted by querying research databases of digital libraries such as \textit{IEEEXplore}, \textit{ACM Digital Library}, and \textit{Elsevier ScienceDirect}, as well as accessible web search engines and websites such as \textit{Google Scholar} and \textit{ResearchGate}. We used search keywords as the ones from the \textit{Index Terms} of the paper. The filtering result{s} were managed based on the content of each article}. A list of common attacks along with their description and example literature is presented in Table \ref{tab:definitions}. The described attacks and threats are a subset of those existing in security studies \cite{alguliyev2018cyber}, i.e., attacks on sensor devices, actuators, computing components, communications, and feedback. Table \ref{tab:definitions} elaborates on attack methods utilized as part of real-world examples presented in Section \ref{s:attackIncidents}. 
A more comprehensive tree diagram of attacks and threats on IoT and cyber-physical systems (CPS) from literature is shown in Fig. \ref{fig:attackleaf}.

\begin{table*}[t!]
\centering
\begin{adjustwidth}{-.15in}{-.2in} 
\caption{
{Common malicious attacks in the IoT spectrum and their definitions.}}
\begin{tabular}{||c|l|l||}
\hline \hline
\textbf{Attack Vector} &
  \multicolumn{1}{c|}{\textbf{Description}} &
  \multicolumn{1}{c||}{\textbf{Example attacks from literature}} \\ \hline
  
\hypertarget{replay2}{\hyperlink{replay1}{Replay attack}} &
  \begin{tabular}[c]{@{}l@{}}An adversary eavesdrops a communication,\\ records a message,  and resends it to the\\ legitimate receiver.\end{tabular} &
  \begin{tabular}[c]{@{}l@{}}Replay attack {\cite{SENGUPTA2020102481}}, relay attack (forwards message\\ without recording it) {\cite{8887286}}, reflection: attacking the\\ challenge response part of the communication.\end{tabular} \\ \hline
\hypertarget{mitm2}{\hyperlink{mitm1}{Man-in-the-middle}} & \begin{tabular}[c]{@{}l@{}} Getting in the middle of a communication\\ between two parties who believe that they\\ are directly communicating with each other. \end{tabular}
  &
  SSL hijacking, HTTPS spoofing, man-in-the-cloud. \\ \hline
\begin{tabular}[c]{@{}c@{}}\hypertarget{cryptanalysis2}{\hyperlink{cryptanalysis1}{Cryptanalysis}}\\attacks\end{tabular} &
  \begin{tabular}[c]{@{}l@{}}The process of analyzing available\\ data to decrypt a message. \end{tabular} &
  \begin{tabular}[c]{@{}l@{}}Ciphertext only-, known plaintext-\\ chosen ciphertext-, chosen plaintext- attacks {\cite{PoC_Cry}}.\end{tabular} \\ \hline
\begin{tabular}[c]{@{}c@{}}\hypertarget{SidechannelAttacks2}{\hyperlink{SidechannelAttacks1}{Side channel}}\\ attacks\end{tabular} &
  \begin{tabular}[c]{@{}l@{}}Attacks based on data gained from the\\ implementation of a system, e.g.,\\ electromagnetic field, power consumption.\end{tabular} & \begin{tabular}[c]{@{}l@{}}
  Cache timing side channel attack,\\ emission security attacks {\cite{8887286}}. \end{tabular} \\ \hline
\begin{tabular}[c]{@{}c@{}}\hypertarget{hwtrojan2}{\hyperlink{hwtrojan1}{Hardware trojans}} \end{tabular} &
  \begin{tabular}[c]{@{}l@{}} Malicious modifications to the\\ design of integrated circuits. \end{tabular} &
  Combinational trojans, sequential trojans {\cite{tsoutsos2014advanced}}. \\ \hline
\hypertarget{covert2}{\hyperlink{covert1}{Covert attacks}} &
  \begin{tabular}[c]{@{}l@{}}The attacker alternates the input of a system\\ and at the same time manipulates the output\\ keeping attack effects undetected.\end{tabular} &
  \begin{tabular}[c]{@{}l@{}}Hardware storage value/transition-based, hardware\\timing value-based, OS storage transition-based, OS\\timing value-based \text{\cite{5654967}}, e.g., Spectre and Meltdown.
  \end{tabular} \\ \hline
\hypertarget{sleep2}{\hyperlink{sleep1}{Sleep deprivation}} &
  \begin{tabular}[c]{@{}l@{}}The rapid exhaustion of the device's power\\ source (e.g., battery) by forcing it to operate\\ an action making it infeasible to enter\\ power-saving mode.\end{tabular} &
  Sleep deprivation {\cite{9219584, 8887286}}, or sleep denial {\cite{SENGUPTA2020102481}}. \\\hline
\hypertarget{clck2}{\hyperlink{clck1}{Clock skewing}} &
  \begin{tabular}[c]{@{}l@{}}The clock skew is compromised by a slight\\adjustment of the timestamps making the\\time sequence change speed causing errors\\in timestamp-based applications.\end{tabular} &
  Suspension, fabrication, masquerade {\cite{skew}}. \\ \hline
\hypertarget{spoofing2}{\hyperlink{spoofing1}{Spoofing}} &
  \begin{tabular}[c]{@{}l@{}}When an entity/program deceives someone\\ else about its identity.\end{tabular} &
  \begin{tabular}[c]{@{}l@{}}RFID spoofing {\cite{SENGUPTA2020102481}}, network spoofing {\cite{7804660}},\\ GPS spoofing {\cite{8221708}}, ARP spoofing \cite{8791653}.\end{tabular} \\ \hline
\begin{tabular}[c]{@{}c@{}}Attacks targeting\\ system reputation\end{tabular} &
  \begin{tabular}[c]{@{}l@{}}A node on the system gets physically\\ tampered or inserted to act maliciously.\end{tabular} &
  \begin{tabular}[c]{@{}l@{}}Device \hypertarget{tamper2}{\hyperlink{tamper1}{tampering}} {\cite{7804660,8796409,9219584}}, \hypertarget{sybil2}{\hyperlink{sybil1}{sybil}}: node operates \\ multiple identities {\cite{9219584,SENGUPTA2020102481}}, fake node injection {\cite{SENGUPTA2020102481},} \\ \hypertarget{wormhole2}{\hyperlink{wormhole1}{wormhole}} (a node sniffs packets at one point of the\\ network and replays them at another point) \cite{SENGUPTA2020102481}.\end{tabular} \\ \hline
\hypertarget{malware2}{\hyperlink{malware1}{Malware}} &
  Malicious software. &
  \begin{tabular}[c]{@{}l@{}} Worm: replicates itself to spread, Adware: throws\\ adds, Virus: replicates itself, modifies software, injects\\ code,  Trojan: disguises as genuine program {\cite{SENGUPTA2020102481}}.\end{tabular} \\ \hline
\begin{tabular}[c]{@{}c@{}}\hypertarget{data2}{\hyperlink{data1}{Data and command}}\\ integrity attack\end{tabular} &
  \begin{tabular}[c]{@{}l@{}}Alternates, creates, and  manipulates false\\ data to a system, network, packet.\end{tabular} &
  \begin{tabular}[c]{@{}l@{}}Data alteration {\cite{7804660}}, FDIA {\cite{SENGUPTA2020102481}}, packet injection,\\ command manipulation {\cite{8887286}}, routing attacks {\cite{8796409}}.\end{tabular} \\ \hline
\begin{tabular}{@{}c@{}}\hypertarget{dos1}{\hyperlink{dos2}{Denial-of-Service}}\\(DoS) \end{tabular}&
  \begin{tabular}[c]{@{}l@{}}Making users unable to access the\\ resources of a device, network, system. \end{tabular} &
  \begin{tabular}[c]{@{}l@{}}Distributed DoS (DDoS) {\cite{7804660,SENGUPTA2020102481}}, flooding {\cite{7804660}},\\ sinkhole {\cite{SENGUPTA2020102481}}, black/grey hole {\cite{8887286}}, packet \\ drop/redirection {\cite{8796409}}, Permanent DoS (PDoS) {\cite{SENGUPTA2020102481}}, \\ channel/radio frequency jamming {\cite{9219584,8887286,SENGUPTA2020102481}}.\end{tabular} \\ \hline
\hypertarget{eave2}{\hyperlink{eave1}{Eavesdropping}} &
  \begin{tabular}[c]{@{}l@{}} Theft of information as it is transmitted\\ over a network, device, etc. \end{tabular} &
  \begin{tabular}[c]{@{}l@{}}Eavesdropping {\cite{7804660,8796409}}, traffic analysis,  spyware {\cite{SENGUPTA2020102481}}.\end{tabular} \\ \hline
  
  \begin{tabular}[c]{@{}c@{}}Command and control\\ attacks (C\&C or C2)\end{tabular}
 &
  \begin{tabular}[c]{@{}l@{}}An attacker controls a computer/device by\\ sending commands to it {\cite{c2}}.\end{tabular} &
  \hypertarget{minject2}{\hyperlink{minject1}{Malware injection}} {\cite{8887286}}, botnet, command injection. \\ \hline
  
Social engineering &
  \begin{tabular}[c]{@{}l@{}}The attacker deceives an authorized user\\ to provide classified information.\end{tabular} &
  \hypertarget{phishing2}{\hyperlink{phishing1}{Phishing}} {\cite{9219584}}, spear phishing, baiting, scareware. \\ \hline \hline
  
\end{tabular}
\label{tab:definitions}
\end{adjustwidth}
\end{table*}

\subsection{Related Work} \label{s:related}

    Different attack categorizations within the IoT landscape exist in literature. For instance, there exist works demonstrating how to leverage the attack layer (e.g, application, network, physical, etc.) \cite{9219584}, the attacker behavior \cite{7804660}, or both \cite{8796409}, in order to construct IoT attack classifications. In the following paragraphs, we provide a short survey of the existing literature discussing security issues within IoT and and various attack taxonomies.

    In \cite{8711834}, the authors review the existing strategies for IoT security and privacy, and utilize their taxonomy to elaborate various concerns and solutions. Their classification is decomposed into three layers, the \textit{information, connectivity} and \textit{application} layers. Each one of these three layers is comprised of two sub-parts, namely the \textit{sensors} and \textit{software}, \textit{connectivity} and \textit{software}, \textit{people} and \textit{connectivity}, correspondingly. Taking into consideration the various technologies involved in each layer, the paper presents the security goals and discusses concerns which could potentially adversely impact such security objectives.
    
    Considering objects characteristics as the different factors between IoT applications,  the authors in \cite{8356843} present a taxonomy where privacy and security are the main focal points. The object characteristics are \textit{automation}, \textit{intelligence}, \textit{storage}, and \textit{processing}. The taxonomy is  segregated into three dimensions, where each dimension analyzes object characteristics, privacy, or security concerns and emphasizes the inter-dependencies and interrelations between them.
    
    In \cite{7804660}, the authors present an IoT attack taxonomy considering attacks from different application fields. 
    The three domains being discussed are smart homes, healthcare, and transportation. The presented comparative attack analysis includes attack targets, weaknesses, and techniques. The analysis investigates attacks which target the exchanged data between nodes or their routing intelligence, data theft through fake websites, and attacks compromising the network infrastructure leading to abnormal or unresponsive states. The attack taxonomy consists of eight categories and aims to enhance security awareness during IoT development as well as provide security analysts with a risk knowledge-base. 
    
    The \textit{device property} category  reflects the attack impact and is divided into the \textit{low-end} class, where low power devices act as the adversary (e.g., smartwatches), and the \textit{high-end} class which includes powerful devices (e.g., laptops) attacking the IoT systems. The second attack category considers the \textit{access level} of the attack. It  can either be \textit{passive} (e.g., \hypertarget{eave1}{\hyperlink{eave2}{eavesdropping}}) or active (e.g, \hypertarget{replay1}{\hyperlink{replay2}{replay attacks}}). 
    The \textit{adversary location} and \textit{attack strategy} categories consider an insider or outsider attacker, and whether the attack will impede the physical or logical functioning of the devices, respectively. The \textit{information damage level} category relates to the data integrity of the measurements of  IoT devices. 
    Based on how the attacker can exploit this information (e.g., interruption, eavesdropping, alteration, fabrication, message replay, \hypertarget{mitm1}{\hyperlink{mitm2}{man-in-the-middle}} attacks, etc.), the corresponding damage on the system can be evaluated. \textit{Host-based attacks} examine the identity that compromises the system, which could either be a user, 
    a vulnerable software, or a hardware-based compromise. 
    The last two attack categories of this taxonomy focus on protocols: 
    \textit{protocol-based} attacks include protocol-related exploitations (e.g., adversaries acting as genuine users and  causing abnormal protocol operations while targeting the availability of the network devices) from  insider and outsider attackers,  
    and \textit{communication protocol stack} which includes attacks targeting the TCP/IP layers (e.g., physical~$\rightarrow$~jamming, data link~$\rightarrow$~collision, network~$\rightarrow$~spoofing, transport~$\rightarrow$~flooding, application~$\rightarrow$~\hypertarget{clck1}{\hyperlink{clck2}{clock skewing}}).

    A brief overview of security risks in IoT and potential mitigation countermeasures is presented in \cite{8796409}. The authors provide a taxonomy of the security requirements in IoT along with a taxonomy based on the vulnerabilities and potential attacks targeting the communication layers. Additionally, the paper investigates IoT security mechanisms encountered in commercial applications of various protocols (i.e., ZigBee, BLE, 6LoWPAN, and LoRaWAN) and analyzes the security posture  of the IoT devices integrating them.  
    The attack taxonomy consists of seven categories based on the communication architecture in IoT as described in \cite{platformsAndProtocols}, and the exploitations used by attackers based on \cite{7467343}. The \textit{edge layer} category includes attacks such as  \hypertarget{SidechannelAttacks1}{\hyperlink{SidechannelAttacks2}{side channel attacks}},
    \hypertarget{hwtrojan1}{\hyperlink{hwtrojan2}{hardware trojans}}, and \hypertarget{dos1}{\hyperlink{dos2}{denial-of-service (DoS)}} attacks. The objective of attacks in this category is to either jam the communication channels using radio waves and device packages \hypertarget{tamper1}{\hyperlink{tamper2}{tampering}}, or to drain the power source of IoT devices by forcing them to be constantly active (e.g., discharge the batteries of IoT nodes). 
    The \textit{access/middle layer} category includes network attacks such as eavesdropping, packet injection, routing attacks, \hypertarget{spoofing1}{\hyperlink{spoofing2}{spoofing}} attacks, packet-redirection and packet-drop attacks. \textit{Application layer} attacks target the software of IoT  devices (e.g., authentication routines, adversarial examples on machine learning algorithms 
    \cite{adversarialMachineLearning}, etc.).
    
    The work in \cite{8796409}  considers also attacker exploitation paths and distinguishes them in four categories. The \textit{ignoring the functionality} category assumes an adversary being able to exploit IoT device features to connect to the local area network (LAN).  
    \textit{Reducing the functionality} attacks aim to block or limit the device functionalities, while 
    \textit{misusing the functionality} attacks opt to operate the device in an incorrect or unauthorized way. 
    The last category describes \textit{extending the functionality} attacks, where the IoT device, despite its intended operational objective, is also used to perform an alternative task, e.g.,  an alarm sensor being also used as a tracking device disclosing the victims' location. Furthermore, the authors provide a qualitative comparison between various IoT devices and their corresponding employed technologies in terms of information, access and functional levels.

    In \cite{9219584}, a taxonomy is proposed categorizing the threats of IoT systems while mapping which security goals are being targeted by each attack. 
    The taxonomy of the attacks is based on the IoT layer, the violated security goals, as well as the knowledge and capabilities of the attacker. The categorization of attacks consists of \textit{cyber-physical}, \textit{middleware}, and \textit{application} attacks, mapping availability, authenticity, accountability, integrity, confidentiality, and access control as the violated goals of any attack.  
    The cyber-physical layer represents the various sensors and actuators deployed in a system and interacting in real-time with the 'outside world' (e.g., other IoT devices) while making decisions and operating  autonomously. Thus, \textit{cyber-physical} category attacks could target both the hardware and the software of a device. Attacks in this category include, among others, \hypertarget{sleep1}{\hyperlink{sleep2}{sleep deprivation}}, physical attacks where an adversary damages the actual device, jamming and \hypertarget{covert1}{\hyperlink{covert2}{covert}} attacks. The middleware serves as the connecting link between the cyber-physical and application layer. For instance, the middleware  simplifies the exchange of information between IoT devices and supports interconnections between devices manufactured by different vendors. Attacks in the \textit{middleware} category target the transported information in terms of routing, addressing and data manipulation, as well as the mechanisms operating in this layer. Example attacks are \hypertarget{sybil1}{\hyperlink{sybil2}{sybil}}, DoS, sinkhole, and \hypertarget{cryptanalysis1}{\hyperlink{cryptanalysis2}{cryptanalysis}}-type of  attacks. 
    The last layer is the application layer which collects and processes data from the physical-layer while provisioning various system workflows (e.g., data processing, graphical interfaces). \textit{Application} layer attacks, such as \hypertarget{phishing1}{\hyperlink{phishing2}{phishing}}, DoS, malicious updates, cryptanalysis and privilege escalation attacks aim to disrupt system services.

    A comprehensive survey of false data injection attacks (FDIAs) detection algorithms is presented in \cite{8887286}. FDIAs are attacks which manipulate the system measurements -- in a congruous way -- to adversely impact the system operation while avoiding triggering any intrusion detection mechanisms. The authors classify various attacks according to the targeted layer and the way they were performed, and conclude with their  proposed taxonomy. The taxonomy consists of four categories.  \textit{Cyber-based} attacks  refer to incidents occurring on the cyber-layer  such as code manipulation, command manipulation, \hypertarget{minject1}{\hyperlink{minject2}{malware injection}}, FDIAs, sleep deprivation, etc. \textit{Network-based} attacks are materialized exploiting virtual network access to a system, without affecting the software, firmware or physical communication link. Such attacks include, among others, DoS, black/grey hole, FDIAs, etc. In  \textit{communication-based} attacks,  an adversary targets directly the physical layer of the communication either by damaging it or by modifying the exchanged information. Such attacks could be GPS spoofing, relay attacks, FDIAs, and channel jamming. The last attack category describes \textit{physical-based} attacks which target the physical integrity of various devices either via tampering or damaging. Other  attacks within this class exploit electromagnetic waves or target the emanations of the system (e.g., light, sound heat, etc.); referred also as emission security attacks. Although the paper demonstrates a taxonomy for CPS-based attacks, it is still relevant and applicable for IoT architectures due to the similarities encountered between CPS and the IoT sector, especially in terms of security \cite{ly2016security}.

    In \cite{SENGUPTA2020102481}, an IoT attack classification is presented.  For each attack category, the authors discuss the corresponding vulnerabilities and countermeasures as well as real-world attack examples. 
    Moreover, the authors present how blockchain technologies are employed in IoT and IIoT environments provisioning security  solutions for different  sectors, i.e., IoT~$\rightarrow$~healthcare, IoT~$\rightarrow$vehicular ad-hoc networks~(VANET), IIoT~$\rightarrow$~supply chain, IIoT~$\rightarrow$~smart grid. An IoT/IIoT security taxonomy  
    is demonstrated along with traditional and blockchain-based security solutions. The domain of \textit{physical} attacks is comprised by attacks where the adversary can have physical access to the IoT system. Attacks under this category are tampering attacks, malicious code injections, radio frequency interference/jamming, fake node injections, sleep denial attack (or sleep deprivation), side channel attacks, and permanent DoS (PDoS) \cite{pdos}.
    \textit{Network} attacks occur when the IoT network is manipulated inhibiting the IoT system's nominal operation. Typical network attacks are RFID spoofing \cite{8065757}, RFID unauthorized access, routing information attacks, selected forwarding \cite{7991968}, sinkhole, \hypertarget{wormhole1}{\hyperlink{wormhole2}{wormhole}}, sybil, man-in-the-middle, replay, DoS, and distributed DoS (DDoS) attacks. In \textit{software} attacks, an adversary exploits the associated software or security vulnerabilities of IoT systems such as \hypertarget{malware1}{\hyperlink{malware2}{viruses, worms, trojan horses, spyware, adware and malware}}. The last category considers \textit{data} attacks which target the data exchanged in IoT systems between the cloud infrastructure, servers, and databases serving as computation storage resources. Data attacks are performed through \hypertarget{data1}{\hyperlink{data2}{data inconsistencies}}, unauthorized access, or data breaches. The attack classification is designed to help IoT/IIoT security analysts determine potential attacks in their respective domain. Additionally, the demonstration of blockchain solutions emerge as efficient ways to overcome the discussed attacks.

Evidently, a great variety of attack taxonomies within IoT ecosystems already exists in the literature. However, each one of the related works investigates IoT attacks under the prism of different security/attack objectives and utilizes diverse criteria for the attack classifications. Some attack classes are extensively discussed and supported by multiple attack paradigms, while others are vaguely described mainly focusing on the attack's target without providing realistic attack examples (e.g., information damage level, host-based) \cite{7804660}. Despite existing work providing real-world examples of attack incidents \cite{SENGUPTA2020102481, 8887286, 8796409}, such papers fail to delineate the attacks in a meticulous way, report in which category these incidents should be allocated, and justify the reasons for such classifications. 
In our taxonomy, we account for the fact that these incidents could not be sufficiently categorized into a single attack class. Notably, adversarial paths overlapping various categories  are common and in many cases different attack paths exist that can induce the same impact. Table \ref{tab:related} summarizes the taxonomies from related work along with the one we present in this paper. In order to illustrate the practicality of our attack taxonomy, in Section \ref{s:attackIncidents}, we discuss real-world attack incidents and map them accordingly. Such real-world attack incidents exemplify the similarities between attacks existing in different IoT deployments (e.g., consumer, commercial, industrial) but also highlight the disastrous effects that some attacks can have in certain operation field contexts stressing the fact that IoT security should be an essential requirement.

\begin{table*}[t]
\centering
\caption{
{Summary of taxonomies.}}
\renewcommand{\arraystretch}{1}
\tabcolsep=0.09cm
\begin{center}
\begin{tabular}{||c|c|c||}
\hline
\hline
\multirow{2}{*}{\parbox{0.9cm}{\centering {{{\bf Paper}}}}} & \multirow{2}{*}{\parbox{7.2cm}{\centering {{{\bf Categories}}}}}  & \multirow{2}{*}{\parbox{9.3cm}{\centering {\bf Details\\}}}  \\
{}       &       {}   &       {}\\ \hline
\multirow{4}{*}{\parbox{0.9cm}{\centering {{{{\cite{7804660}}}}}}} &\multirow{4}{*}{\parbox{7.2cm}{\centering {{{{Device Property, Access Level, Adversary, Location, Attacks Strategy, Information Damage Level, Host-based attacks, Communication Stack Protocol.}}}}}}&\multirow{4}{*}{\parbox{9.3cm}{\centering {{{{Each category provides a subcategorization of attacks which often overlap with each other, e.g., access level overlaps with  the information damage level. Examples are provided, but without  mapping to real-world incidents.}}}}}}\\
{}       &       {}   &       {}\\ 
{}       &       {}   &       {}\\ 
{}       &       {}   &       {}\\ \hline
\multirow{4}{*}{\parbox{0.9cm}{\centering {{{{\cite{8796409}}}}}}} &\multirow{4}{*}{\parbox{7.2cm}{\centering {{{{Edge Layer, Access/Middleware Layer, Application Layer, Ignoring the Functionality, Reducing the Functionality, Misusing the Functionality, Extending the Functionality.}}}}}}&\multirow{4}{*}{\parbox{9.3cm}{\centering {{{{It provides security mechanisms adopted by communication protocols, however, combining communication layer and attacker behavior could cause uncertainty in classification. 
Examples are provided, without  mapping to real-world incidents.}}}}}}\\
{}       &       {}   &       {}\\ 
{}       &       {}   &       {}\\ 
{}       &       {}   &       {}\\ \hline
\multirow{4}{*}{\parbox{0.9cm}{\centering {{{{\cite{9219584}}}}}}} &\multirow{4}{*}{\parbox{7.2cm}{\centering {{{{Cyber-physical layer attacks, Middleware layer attacks, Application layer attacks.}}}}}}&\multirow{4}{*}{\parbox{9.3cm}{\centering {{{{The work develops simply a naming scheme used in a taxonomy in order to correlate it with violated security goals. Examples are provided, but without  mapping to real-world incidents.}}}}}}\\
{}       &       {}   &       {}\\ 
{}       &       {}   &       {}\\ 
{}       &       {}   &       {}\\ \hline
\multirow{4}{*}{\parbox{0.9cm}{\centering {{{{\cite{8887286}}}}}}} &\multirow{4}{*}{\parbox{7.2cm}{\centering {{{{Communication-based, Cyber-based, Network-based, Physical-based.}}}}}}&\multirow{4}{*}{\parbox{9.3cm}{\centering {{{{The taxonomy emphasizes in cyber-physical and data integrity attacks, and lacks specific IoT attacks and context. Examples are given but emphasized in false data injection attacks. Incidents are discussed but not mapped to the taxonomy.}}}}}}\\
{}       &       {}   &       {}\\ 
{}       &       {}   &       {}\\ 
{}       &       {}   &       {}\\ \hline
\multirow{4}{*}{\parbox{0.9cm}{\centering {{{{\cite{SENGUPTA2020102481}}}}}}} &\multirow{4}{*}{\parbox{7.2cm}{\centering {{{{Physical Attacks, Network Attacks, Software Attacks, Data Attacks.}}}}}}&\multirow{4}{*}{\parbox{9.3cm}{\centering {{{{A well-structured taxonomy with countermeasures and real-world examples. However, no comprehensive description is provided  to the mapping of incidents.}}}}}}\\
{}       &       {}   &       {}\\ 
{}       &       {}   &       {}\\ 
{}       &       {}   &       {}\\ \hline
\multirow{4}{*}{\parbox{1.4cm}{\centering {{{{Proposed Taxonomy}}}}}} &\multirow{4}{*}{\parbox{7.2cm}{\centering {{{{{Device, Infrastructure, Communication, Service.}}}}}}}&\multirow{4}{*}{\parbox{9.3cm}{\centering {{{{We use the pillars of consumer, commercial, and industrial IoT along with real-world cases to map them with our taxonomy. Case studies are discussed for every category and potential mitigations are provided. 
}}}}}}\\
{}       &       {}   &       {}\\ 
{}       &       {}   &       {}\\ 
{}       &       {}   &       {}\\ \hline \hline
\end{tabular}
\label{tab:related}
\end{center}
\end{table*}

\subsection{Proposed Taxonomy} \label{s:ProposedTax}
        
    Factoring the related work literature, we propose a taxonomy and leverage it to investigate attacks that threaten consumer, commercial, and industrial IoT ecosystems. We discuss IoT attacks which target the aforementioned three IoT domains in order to demonstrate the applicability of our taxonomy even for diverse attack scenarios targeting different system assets and aiming to compromise distinct security objectives.
 
    \begin{figure*}[t]
        \centering
            \includegraphics[width=0.85\linewidth,trim={0cm 0cm 0cm 0cm}]{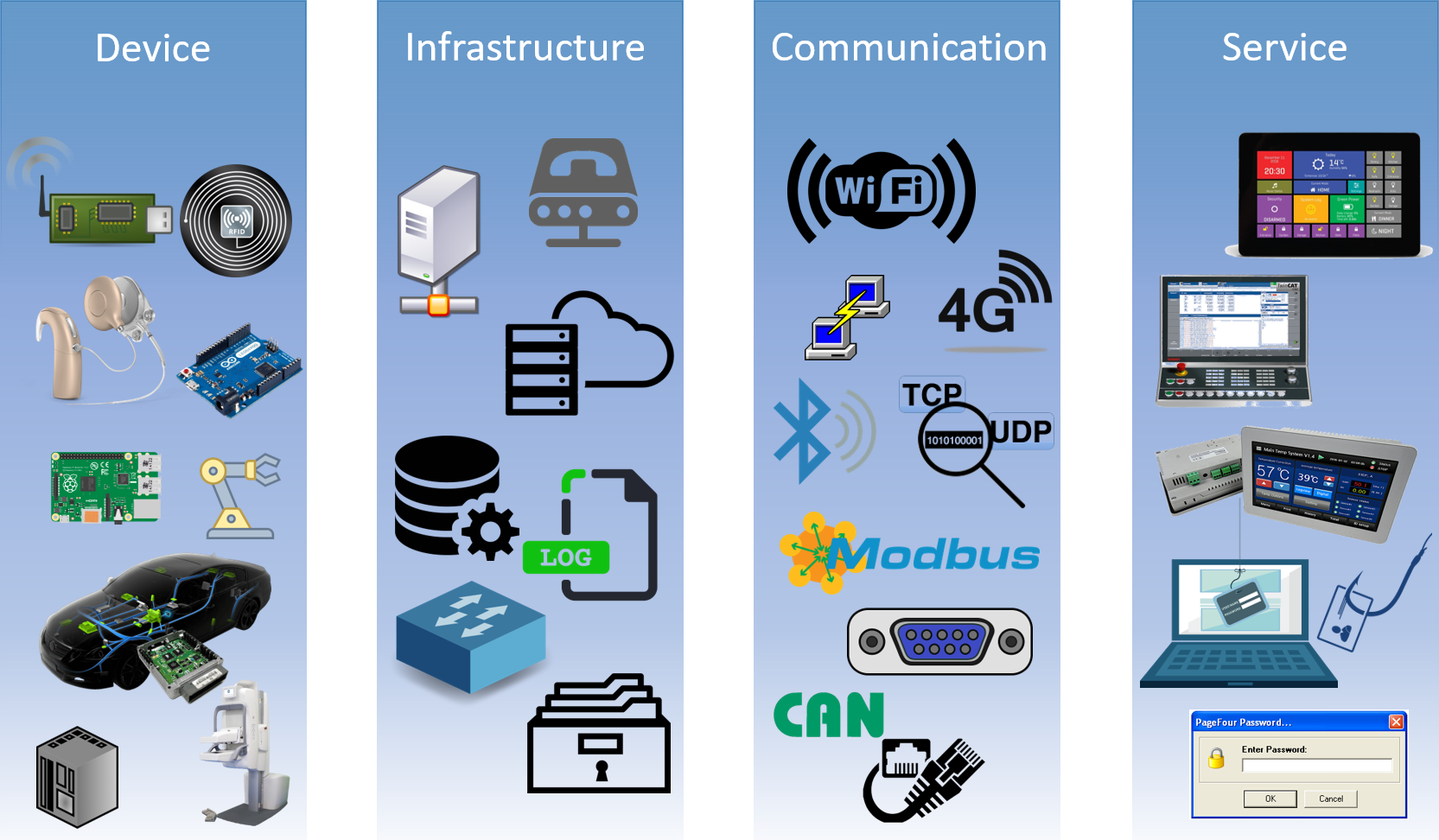}
            \caption{{Attack taxonomy and corresponding components.} }
            \label{fig:attack_taxonomy}
    \end{figure*}   
    
     {Due to the ubiquity of IoT devices spanning into multiple fields, we classify IoT attacks into four categories: \textit{Device},  \textit{Infrastructure}, \textit{Communication} and \textit{Service} attacks. We opt for this classification since it provides a streamlined approach to categorize attacks in an aggregate way, overcoming the requirement to create different attack classes based on the adversarial target, compromised security objective, attacker tactics, etc.  These classes are also supported by the presented real-world cyberattacks in Section \ref{s:attackIncidents}.} Some attacks can fall under more than one categories depending on their threat model as well as the means they require in order to be accomplished.  Malicious code injection attacks,  for example, could be considered a \textit{Service} attack when performed on a website, whereas if performed through a serial port of an IoT node could be regarded as a \textit{Device}-type of  attack. Similarly, a DoS attack may flood a \textit{Communication} channel rendering data exchanges infeasible. At the same time, it could also target the \textit{Infrastructure} of an IoT system by abusing the host's available resources (e.g., CPU or memory capacity) \cite{ddos}.
     
     The \textit{Device} category consists of attacks where an adversary aims to damage or tamper with the hardware components or the ``things'' of an IoT system. This type of attacks can also be referred to as perception layer attack \cite{aarika2020perception}, containing sensors and actuators as well as edge computing devices that could be micro-controllers, programmable logic controllers (PLCs), smartwatches, medical implants, etc. Such devices capture data from their environment physically or digitally, convert them to digital signals, and forward them to other IoT components or infrastructure through the communications channel. Attacks on the hardware and software parts of the device, either with or without physical access to the IoT device, would cause an abnormal operating condition of the IoT system, affecting the exchange of information within the device environment.  
     Such device-type of attacks could be performed, for instance, through the exploitation of hardware ports, as is the case in node/device tampering \cite{nuke_mahajan_thool_2013}, through malicious code injection  leading to system dysfunction \cite{u.farooq_waseem_khairi_mazhar_2015},  hardware trojans \cite{tsoutsos2014advanced}, node or object jamming \cite{nuke_mahajan_thool_2013}, sleep deprivation attacks \cite{bhattasali_chaki_sanyal_2012}, and remote firmware update attacks \cite{8966174, kuruvila2020hardware, 7436314}.
     
     {The second category of our attack taxonomy includes \textit{Infrastructure} attacks {that} target the ``back-end'' of a system which is the data access layer, including data storage and data processing. The data access layer is  connected with the database management which could be local or at the cloud. Edge computing includes packet content inspection, data filtering, cleanup, and aggregation mechanisms. As a result, events are generated, filtered, compared, joined in complex processes, evaluated, and aggregated. The information is then integrated according to an interface employed by the system applications. Most infrastructure attacks threaten either the physical integrity or availability of data or devices located at the edge of the network (e.g., jamming attacks). For example, compromising sensitive data through weak passwords or social engineering could lead to account hijacking attacks \cite{5461732}, and as a consequence to data manipulation, corruption, insertion, loss, 
     or scavenging \cite{5719001}}.

     {\textit{Communication} attacks impact the broadcast and exchange of information between IoT devices. Communication attacks can target  communication protocols, communication standards, communication technology, or even the communication channel which could be wireless or physical. Moreover, this category also considers attacks endangering the switching, routing, and data exchange  at the network layer, protocol implementations as well as translations between different protocols. A high number of the reported IoT attacks in literature belong to this category due to the exposed  communication channels and  lack of sophisticated security functionalities. IoT communications can be vulnerable against a variety of attacks such as  traffic analysis \cite{nuke_mahajan_thool_2013}, eavesdropping \cite{._2014}, interception \cite{zhou_yang_yang_2017}, etc. On the network side, an adversary could potentially arp-poison the communication endpoints and pursue a man-in-the-middle attack \cite{7442758}, flood the network with packets causing a DoS or DDoS attack \cite{mahesh_rodrigues_2014}, or even generate network disruptions via routing information attacks. An attacker could also record sensitive information and carry out harmful attacks such as RFID cloning \cite{Mitrokotsa_classificationof}. In such scenarios, attackers record exchanged RFID tag data, and then exploit  them conducting replay attacks.}

    The last category in our taxonomy considers \textit{Service}-type of attacks. By service, we refer to the inherent functionality and services that the system provides to users through various processes. System services include the front-end along with the IoT software that orchestrates several parts of the system in coordination with the end-users. For instance, system reports provided to users belong to the adversarial targets of this attack category. Such reports include data analytic results presented in graphical environments where users can interact and provide input information or issue control commands. 
    Example attacks in this category are phishing attacks, social engineering \cite{7813778}, and control hijcacking, i.e., methods used by adversaries to directly manage IoT applications \cite{papp_ma_buttyan_2015}.   {Most of the attacks in this category are \textit{application layer} attacks, e.g.,  malicious scripts, cryptanalysis attacks, exploitation of buffer overflow vulnerabilities, attempting to obtain sensitive information from the memory of the application, reverse engineering methods decompiling the executable software to find potential exploits, etc.}

    A conceptual diagram of the proposed IoT attack taxonomy, including popular attack targets for each of the attack categories {as referred from our proposed taxonomy}, is depicted in Fig. \ref{fig:attack_taxonomy}. In the following section, we discuss notorious IoT attack incidents derived from the consumer, commercial, and industrial sectors, and explain how they can be investigated and mapped using our taxonomy. Furthermore, we provide detailed analyses regarding the IoT vulnerabilities of each attack incident, the methods that adversaries followed to realize the attack, the resulting system impact following the successful implementation of the attack, and potential mitigation strategies which could have averted these attack cases.
\def\ucr{\scalebox{1.2}{\stackinset{c}{}{c}{-.1pt}{%
  \textcolor{white}{\sffamily\bfseries\tiny ?}}{%
  \rotatebox{45}{$\blacksquare$}}}}
 
\section{Attacks use cases targeting IoT deployments} \label{s:attackIncidents}

In this section, we analyze real-world IoT incidents targeting the consumer, commercial and industrial domains, and describe how these attacks can be classified following our taxonomy. 
{We elaborate on the vulnerabilities of these IoT ecosystems and discuss how adversaries exploit them in order to perform their attacks. We describe the impact of the attacks along with security countermeasures to prevent their proliferation in  IoT systems. Our attack taxonomy aims to assist security analysts design secure and resilient IoT systems exposing the underlined characteristics of IoT attacks.
We select use cases with diverse adversarial objectives as well as attack procedures covering the diverse and immense IoT attack surface while also highlighting the existing system vulnerabilities.}

\subsection{Incidents Targeting Consumer IoT Devices}
Consumer IoT devices exist in various aspects of our everyday lives  due to the plethora of capabilities they feature. Typical examples of consumer IoT devices include smartphones, smart watches, indoor and outdoor cameras, etc.   
In the following subsections, we discuss three consumer IoT attacks covering voice assistant devices, baby monitoring cameras, and botnets affecting consumer products along with the inherent vulnerabilities of each attack scenario, system impact, and potential mitigation plans to overcome such unfavorable situations.

\subsubsection{Case Study 1 -- Voice Assistant }
\label{ch:voiceAssistant}
\begin{figure}[t]
        \centering
            \includegraphics[width=0.9\linewidth,trim={0cm 0cm 0cm 0cm}]{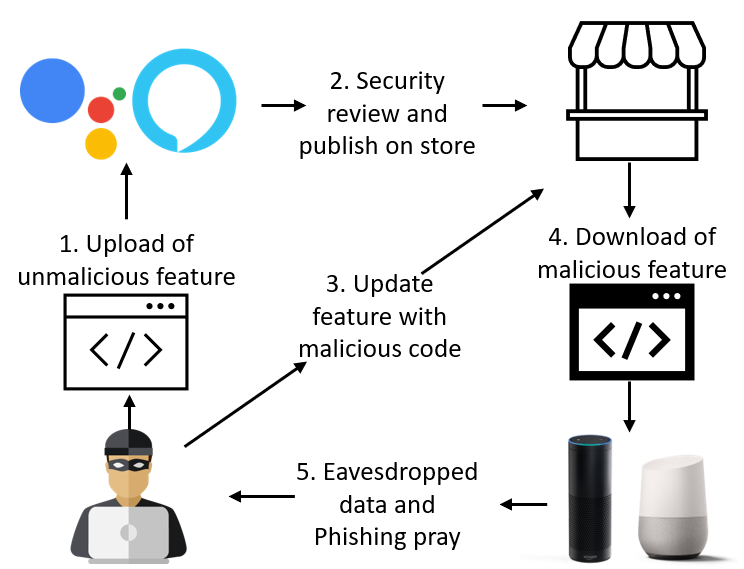}
            \caption{Home assistant attack deployment methodology.}
            \label{fig:home_assistants}
        \end{figure}

    The introduction of voice assistance on smartphones thrusted the emergence of home assistants. However, the wide adoption of IoT home assistants (e.g., Google assistant, Amazon echo, etc.) was accompanied by a new set of vulnerabilities. {For instance, researchers, at the Security Research Labs~(SRL) \cite{braunlein_frerichs_2019}, managed to mount attacks on home assistants exploiting device vulnerabilities, compromising their operation, and intruding user privacy.} The researchers developed specific software frameworks able to leverage implementation bugs and eavesdrop users or plant sophisticated phishing attacks~\cite{charlton_2019}. The vendors name the aforementioned software ``skills'' and ``actions'', for Amazon Echo and Google Assistant, correspondingly. 
    Although ethical hacking approaches, such as  \cite{braunlein_frerichs_2019}, demonstrate the feasibility of these types of attacks, malicious users could also follow similar methods. In Fig.~\ref{fig:home_assistants}, we illustrate the attack methodology targeting smart home assistants. {Attackers, before mounting their attack, must first bypass Google's or Amazon's preliminary security review, once they upload their custom-built applications to the corresponding app-stores.} Once this security check is successfully evaded, adversaries would prompt users to update the newly installed application; however, they need to ensure that this update would not trigger any security mechanism or initiate a secondary application review process (from Google or Amazon). Following the application update, harmful features could be also ported to the assistant devices enabling \textit{Service} attacks by imitating other trustworthy apps, request for security credentials, overhear user conversations, or perform other types of identity-theft related attacks. Apart from \textit{Service} attacks, the loss of user data confidentiality can also lead to \textit{Infrastructure}-aimed attacks, e.g., account hijacking to access and modify confidential user data, or \textit{Communication}-based attacks via spoofing the voice assistance features or accessing the user home network.

    To delude users into thinking that the newly installed application features are not working properly, researchers at SRL strategically changed the welcome message of the named application while demonstrating that the security mechanisms can be evaded. The new message would include a sequence of unpronounceable characters {(e.g., ``\ucr. '' or ``U+D801, dot, space'')}, since this would create an artificial silent delay leading users into believing the application might have encountered some bug. Immediately after the pause, the device would ask the user for the password since ``critical security updates'' needed to be installed in the device. Despite that some users might be skeptical in sharing their security credentials, still many would fall victims to this sophisticated phishing attack. {Following the same procedure, i.e., leveraging the same sequence of characters to create this fake silence effect and using the ``skill'' software framework, the SRL researchers were able to ``selectively wake up'' the voice assistant devices using specific trigger words, e.g., password, to eavesdrop the following communication. This  eavesdropping attack was demonstrated using the Google Home assistant, enabling attackers to receive a comprehensive log of the users' conversations.}

    It is evident that user data privacy is the most important principle for the security and privacy related to consumer electronics products, and the one that seems the most overlooked according to the SRL study. Given how easily attackers -- leveraging voice assistant vulnerabilities -- could steal user credentials or perform account hijacking attacks, the SRL researchers suggest that any new application feature should be extensively reviewed every time it gets updated from the corresponding app-store. {Additional countermeasure suggestions are, provided in detail in~\cite{braunlein_frerichs_2019}, advocating that silent messages, unpronounceable characters, as well as keywords should be handled diligently, impeding adversaries aiming to build exploits from such context-sensitive details.}

\subsubsection{Case Study 2 -- Baby Monitoring Cameras} \label{ch:baby_monitoring}

    Popular consumer electronics devices, especially in families with newborns, are baby monitoring cameras~(BMCs). In 2018, a family in South Carolina realized that their BMC was operating suspiciously; their skepticism was validated since their device was indeed compromised~\cite{domonoske_2018}. The security analysis by SEC Consult led to the identification of a multitude of vulnerabilities for the underlined BMC~\cite{secconsult_2018}. For instance, TCP ports $554$ and $5000$ could be exploited without authentication from anyone with access to the BMC's local network. Although local access is by no means granted apriori, this constraint can be bypassed if unsuspicious users fail to protect their home networks, making every connected device vulnerable to adversaries. Additionally, the BMC furnishes a default-enabled peer-to-peer~(P2P) cloud feature which streams the unencrypted captured video to the cloud, thus further increasing the device threat surface. {Furthermore, the default password, a common practise for consumer IoT electronics, is neither  device-specific nor randomly generated.} Last, the BMC features two universal asynchronous receiver-transmitter~(UART) interfaces which grant complete administrator control, via a root shell, to anybody who can acquire physical access to the device.

    Inspecting the reported vulnerabilities, the BMC is evidently insecure since it can be targeted by any type of attack ranging from the \emph{Device} category, e.g., by exploiting the exposed UART ports, all the way to the \emph{Service} category, e.g.,  via the default-enabled P2P streaming features. The security analysis of the specific device demonstrated that attackers could exploit the BMC's P2P cloud feature to compromise the device and violate users' privacy~\cite{secconsult_2018}. By scanning for valid device IDs with default credentials, adversaries are able to connect, a \textit{Service} category attack, to the camera and spy on victims. In this respect, it can be seen as an \emph{Infrastructure}-based attack~(access to cloud recordings) which could be materialized through traffic analysis and packet eavesdropping.
    
    \begin{figure}[t]
    \centering
        \includegraphics[width=6.5cm,trim={0cm 0cm 0cm 0cm}]{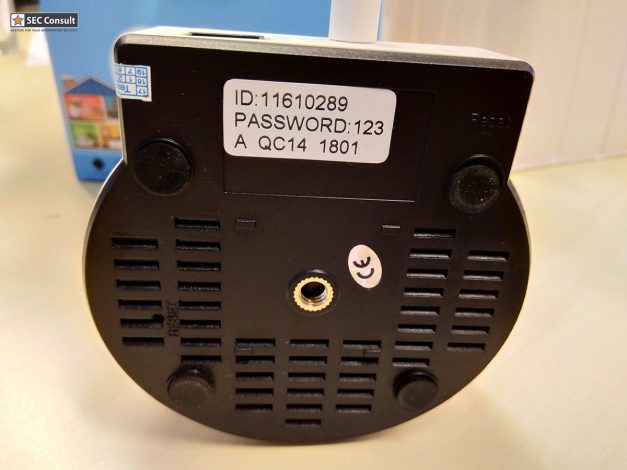}
        \caption{Baby monitoring camera with default credentials \cite{secconsult_2018}.}
        \label{fig:babymonitoring}
    \end{figure}
    Even though the primary target of this event remains the users' privacy, other factors should not be overlooked. The default credentials usage for example as illustrated in Fig. \ref{fig:babymonitoring}, can lead to possible attacks targeting the \emph{Service} aspect of the device as well, locking authorized users out, or providing adversaries with access to customer local networks via the compromised devices. {Researchers, manufacturers, and service providers, in their efforts to safeguard customer security, prompt users to change default credentials, disable features which are not being used, patch their devices with the latest firmware updates, and ensure that encrypted traffic is enabled.} Although these countermeasures cannot deter determined adversaries, they can significantly reduce the potential entry points for attacks.

\subsubsection{Case Study 3 -- The Mirai Botnet} \label{ch:mirai}
    
    Although cyberattacks targeting single consumer IoT devices can have significant consequences to the user's data privacy, as was the case with voice assistants and BMCs, disruptive attacks targeting multiple IoT devices at large have also been reported. We refer to this type of attacks  compromising and controlling multiple connected devices as \emph{botnet attacks}. A well-known example of such attack incident, targeting mainly consumer electronic IoT devices,  is the Mirai botnet, initially discovered by the \textit{MalwareMustDie} non-profit security organization in August 2016 \cite{7971869}. {However, the first generalized Mirai attack occurred in September of the same year, infecting 65,000 devices during the first 20 hours of operation and with a total of 600,000 compromised devices by the end of the Mirai malware's life-cycle (February 2017).} 
    
    The Mirai botnet attack was noticed when  \textit{OVH}, a cloud computing company, and \textit{Krebs on Security}, a blog website investigating cybercrime stories, fell victim to a Distributed Denial-of-Service~(DDoS) attack. The attack renders the enterprise networks unresponsive by overflowing them with artificial traffic saturating the network bandwidth. More specifically, in the Krebs case the traffic peaked at approximately $623$ Gbps. The botnet victims included game servers, anti-DDoS providers, telecommunication firms, political websites, and some suspicious sites hosted in Russia~\cite{themiraibotnet}.

\begin{figure}[t]
        \centering
                \includegraphics[width=8cm,trim={0cm 0cm 0cm 0cm}]{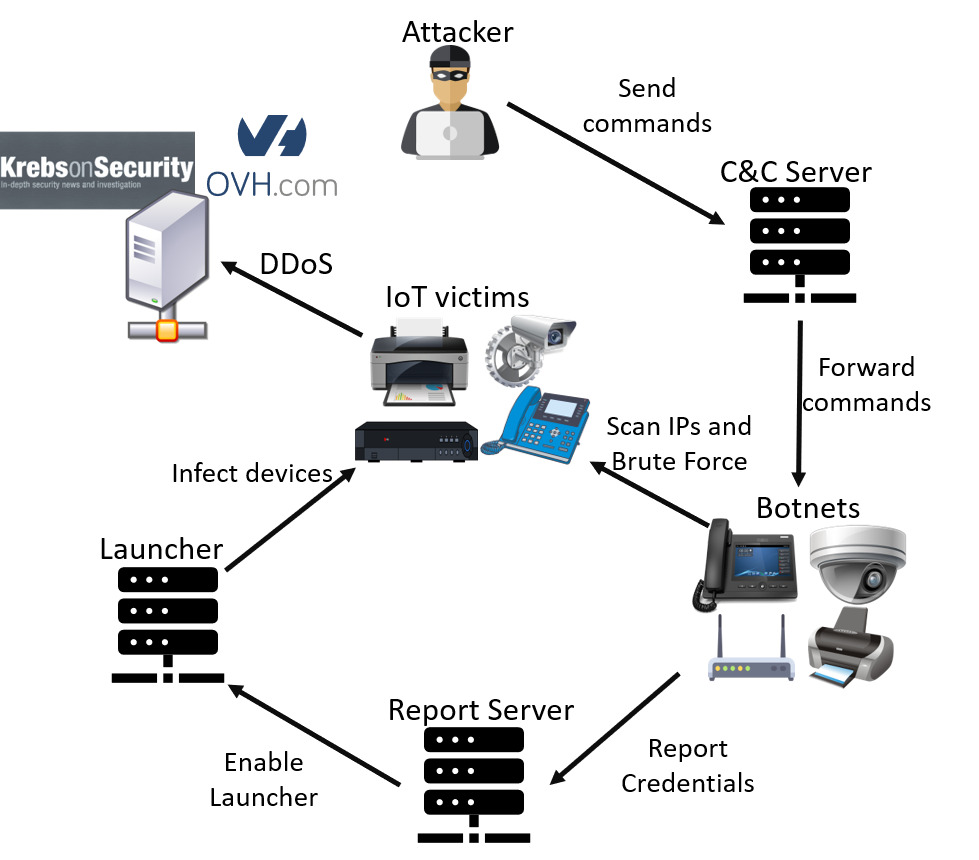}
            \caption{Mirai botnet operation methodology.}
            \label{fig:mirai_botnet_operation}
        \end{figure}

    The Mirai attack procedure includes an initial scan for potential victims using pseudo-random IPv4 network addresses; once a device is found, a \textit{Telnet} or SSH connection is attempted using 10 username and passwords pairs randomly selected from a preconfigured list of $62$ credentials. If the authentication succeeds and the connection is established, the botnet sends the connection credentials to the \textit{report server}, which enables a \textit{loader} which infects the victim with a device-specific malware. The aim of the malware is to remain undetected by antivirus programs, thus it deletes the corresponding downloaded binary and masks the malicious process name to hide its existence. Additionally, it blocks any other processes related with similar infections, or other Mirai variants. The infected device is then remotely managed by the \textit{control server}, enabling the attacker to launch a DDoS attack while scanning for new vulnerable devices concurrently.  
    Fig.~\ref{fig:mirai_botnet_operation} illustrates the Mirai botnet operation, as was prescribed in the source code published by its developers in September 2016, with the initial first victims of the incident. 
    
    As mentioned above, in order to successfully deploy a Mirai botnet attack, a significant number of controlled devices is required. To acquire access to these vulnerable IoT devices, first an IPv4 network address scan is performed. This step, which can be portrayed as \textit{Communication} category attack, targets such devices since they have a much higher probability to use default or non-randomized credentials. With password cracking techniques (e.g., dictionary attacks, rainbow tables, brute-force attacks, etc.) belonging to the \textit{Service} attack class, many IoT devices were successfully compromised. Then the malware binary was installed, handling full device control to the attacker. {Finally, the Mirai botnet can also be seen as an \textit{Infrastructure} type attack, given that the recipients of these DDoS attacks were mainly enterprise networks belonging to the infrastructure fabric.}

    The consumer IoT devices mainly targeted by Mirai were IP cameras, digital video recorder~(DVRs), printers, and routers. This corresponds to an approximate cost of $\$13.50$ per device for consumers, and around $\$4,207.03$ per hour that the network bandwidth is overflown with traffic~\cite{osborne_2018}. The impact is much higher for Domain Name System~(DNS) providers who should have invested in DDoS countermeasures and network resource redundancy. Mirai evolved and diverged into thousands of variants; some of them are more immune to detection mechanisms, others leverage different protocols or dictionaries to deploy the attack. {The Mirai botnet stopped expanding after the arrest of its creators, who, in order to atone for their committed cybercrimes, contributed in the development of the counter-IoT botnet honeypot ``WatchTower''~\cite{wright_2019}.} Regarding the mitigation and avoidance of botnet attacks, the authors of~\cite{themiraibotnet} highlight the importance utilizing randomised password in consumer electronics devices. Furthermore, avoiding default network configurations, and applying Address Space Layout Randomization~(ASLR) -- which prevents exploits of memory vulnerabilities -- can also inhibit such attacks. Automatic updates under secure frameworks are encouraged, alongside notification alerts when suspicious behavior is detected (e.g., unexpected network traffic). Finally, the adoption of immune-to-fragmentation attacks operating systems and by extension withdrawal and upgrade of outdated systems can impede the proliferation of botnet attacks.

\subsection{Incidents Targeting Commercial IoT Devices}

In the commercial sector, the employment of distributed systems -- comprised by interconnected devices which collect, process, and analyze data autonomously before transmitting them to central processing units -- has been widely endorsed in diverse operational scenarios (e.g., smart buildings/cities, transportation, medical facilities, etc.). Along with the commercial IoT infrastructure growth, however, vulnerabilities also appear and being  exploited by malicious adversaries. In this part, we discuss three commercial IoT attacks covering avionic systems, vehicles, and healthcare medical devices. 

\subsubsection{Case Study 4 -- Aircraft Avionics} \label{ch:boeing}

Avionic systems are prominent examples of commercial IoT architectures in which any malicious attack would result in disastrous effects. 
In 2018, an unprotected server was discovered on Boeing's network which allowed download access to the avionics system provider data and code specifically crafted to run on the company's 737 and 787 passenger jets~\cite{greenberg_2019}. The analysis of this publicly accessible information (including reverse engineering the acquired binary files), in addition to the available literature about the network configuration and the IoT infrastructure of the aircraft models, led to uncovering  vulnerabilities within the Boeing 787 aircraft system~\cite{santamantra}.

The security analysis in~\cite{santamantra} reveals references to many insecure code functions on the  Crew Information System - Maintenance System~(CIS/MS) module such as \verb|strcpy|, \verb|sprintf|,  and \verb|strcat|, which can potentially cause \textit{Service}-type of attacks like integer and buffer overflows. Additionally, \textit{Infrastructure} attacks could be initiated since vulnerable execution paths (e.g., unauthorized user commands) could allow for out-of-bound reads or writes as well as memory corruptions. Alarmingly, more serious vulnerabilities were discovered in some of the obtained \verb|.vex| files. Examples include stack and buffer overflows, remote code execution, a vulnerable Trivial File Transfer Protocol (TFTP) server, an insecure system-call handler who could lead to privilege escalation, and Return-Oriented Programming (ROP) exploits. Based on the security analysis findings, various potential scenarios could be formed in which the discovered vulnerabilities (including zero-days) can be exploited leading to destructive attacks. Such scenarios can attain unauthorized access to the Common Data Network~(CDN) which connects most of the aiplanes systems (i.e., through the Common Computing Resource Cabinets to the fuel quantity, low pressure, and lightning systems).

{The IoT and communications network of a Boeing 787 consists of the following components:}
\begin{enumerate}[label=(\roman*)]
    \item The high-integrity CDN  which is considered the backbone of communication for the aircraft facilitating system information exchange between the distributed airplane systems. It is physically connected through fiber-optic and copper links with the Aeronautical Radio, Incorporated~(ARINC) cabinet switches and the Remote Data Concentrators.
    \item The \textit{Common Computing Resource Cabinets~(CCR)}  which include different modules, network switches, and graphic generators
    \item The \textit{Remote Data Concentrators~(RDCs)} which support system analog and discrete signals alongside serial digital interfaces.
    \item The \textit{Core Network Cabinet~(CNC)} which is responsible for the data segregation between the Open Data Network~(ODN), the Isolated Data Network~(IDN), and the CDN.
\end{enumerate}

\begin{figure}[t]
        \centering
                \includegraphics[width=0.9\linewidth,trim={0cm 0cm 0cm 0cm}]{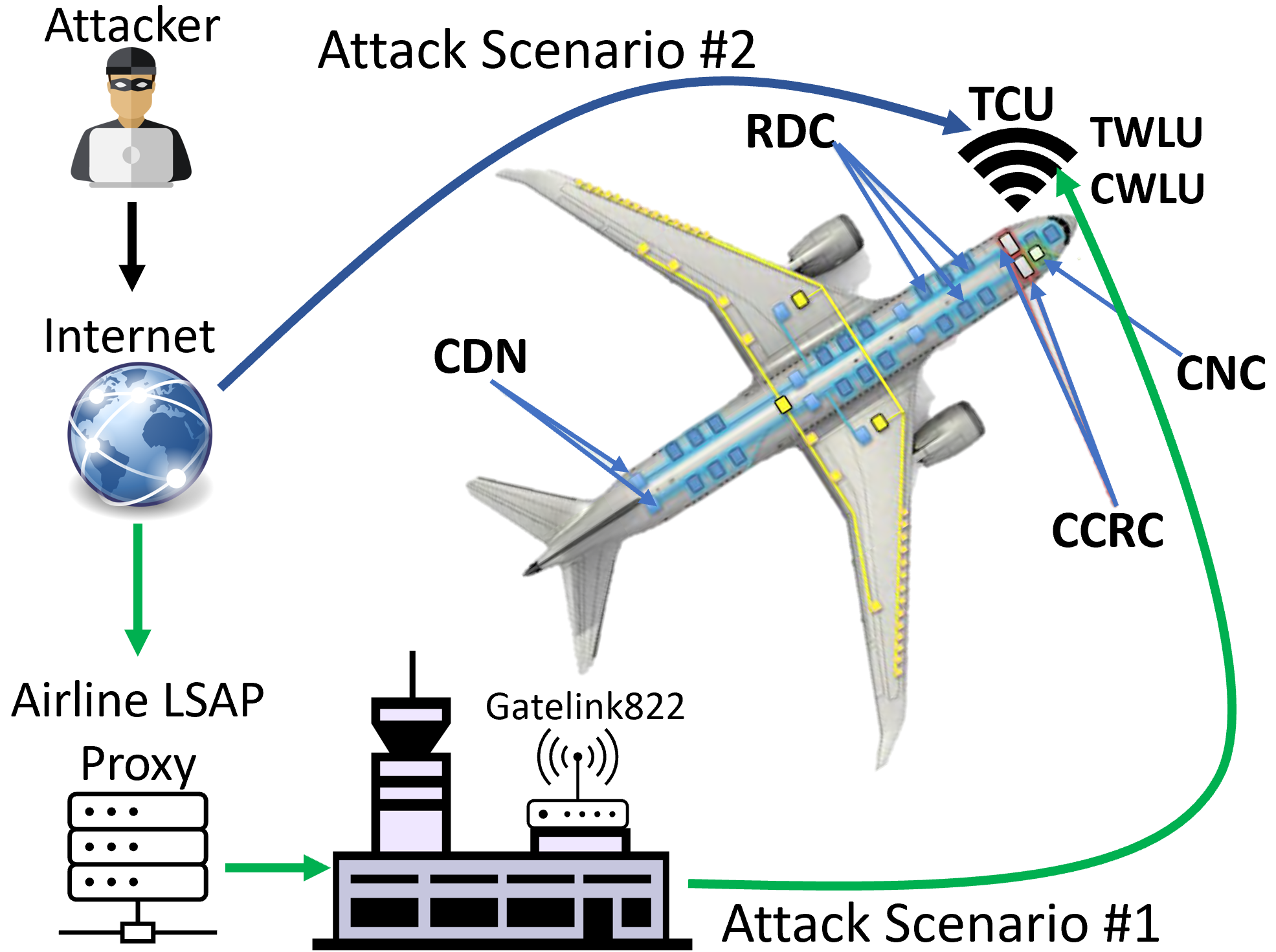}
            \caption{Attack scenarios targeting Boeing 787.}
            \label{fig:boeing 787}
\end{figure}

Leveraging the aforementioned vulnerabilities, two \textit{Communication}-based attack scenarios are depicted in Fig. \ref{fig:boeing 787} targeting the network infrastructure of the 787 aircraft. The first scenario considers an attacker who compromises the Internet-Accessible Loadable Software Aircraft Parts~(LSAP) proxy server, which was discovered that two vulnerable instances of such servers exist (including 737's and 787's data). 
Then the attacker through the proxy can access the data exchanged between the ground control tower and the on-board electronic distribution system of the aircraft~\cite{patent1}, i.e.,  uplink/downlink requests between the ground and the plane.
Another attack path leveraging the LSAP proxy server includes the use of the on-ground aircraft wireless communication Gatelink822 
to access the aircraft's Terminal Wireless LAN Unit~(TWLU) or the Crew Wireless LAN Unit~(CWLU). Then, the attacker can access the IDN, through various Ethernet Gateway Module~(EGM) rules, and compromise the CDN exploiting the pre-referenced vulnerabilities. 
The second scenario 
utilizes the Terminal Cellular Unit (TCU) or satellite communication links which could be exposed to the Internet (via a public IP). According to~\cite{santamantra2}, this hypothesis is likely to be true granting access to the TCU. Once again, through EGM rules, the adversary can reach the CIS/MS and finally jeopardize the CDN.

For the mitigation of the reported attacks, the analysis provided in \cite{santamantra} proposes the use of the x86 32-bit CPU No-Execute (NX/XD) hardware mitigation, supported by the inherent hardware (i.e., Intel Pentium M 32-bit processor). {Furthermore, the adoption of compiler-level mitigation for insecure functions such as} \verb|strcpy|, \verb|sprintf|, etc. are strongly advised to prevent buffer- and stack-based overflow attacks. Finally,  secure firmware  updates -- especially for the mission-critical subsystems, e.g., Bus Power Control Unit~(BPCU), Generator Control Unit~(GCU), Electronic Engine Control~(EEC), Wing Ice Protection System~(WIPS), etc. -- attested by integrity checks and controls need to  verify the authenticity of the firmware, and inhibit the counterfeit firmware impact in the event of an attack. Although Boeing and Honeywell~(the main firmware supplier of Boeing) confirmed the existence of the majority of the aforementioned vulnerabilities, they stated  that such vulnerabilities cannot be exploited and security mechanisms are in place to counter them. Indeed, a catastrophic attack might  not be entirely realized utilizing solely the findings provided in~\cite{santamantra}, however,  {``aircraft security is far from a solved area of cybersecurity research''}~\cite{greenberg_2019}.

\subsubsection{Case Study 5 -- Heavy-Duty Vehicles} \label{ch:heavy_commercial_vehicles}

Similar to consumer vehicles, heavy-duty commercial vehicles such as buses, trucks, trains, and tractors are integrating new technologies and features which rely on information and communication 
interfaces and protocols. Despite such technology integration contributes to safer and more efficient transportation as well as to comfort improvements, it also increases the potential entry points for adversaries~\cite{hackingtrucks}. {In 2015, security researchers -- by exploiting a zero-day vulnerability on the IoT-enabled entertainment system -- demonstrated that vehicle systems, e.g., engine, steering, braking, etc., can be remotely and/or maliciously controlled\cite{9213827}.} Similarly in 2016, researchers from the University of Michigan demonstrated the existence of vulnerabilities by hacking the internal Controller Area Network (CAN) bus of both a truck and a school-bus~\cite{198487}. They compromised these heavy-duty vehicles by physically connecting to their on-board diagnostics (OBD) port, and through it controlled the throttle and engine brakes, as well as the indicators for various gauges (e.g., fuel, brakes, etc.).

\begin{figure}[t]
    \centering
        \includegraphics[width=8.7 cm,trim={0cm 0cm 0cm 0cm}]{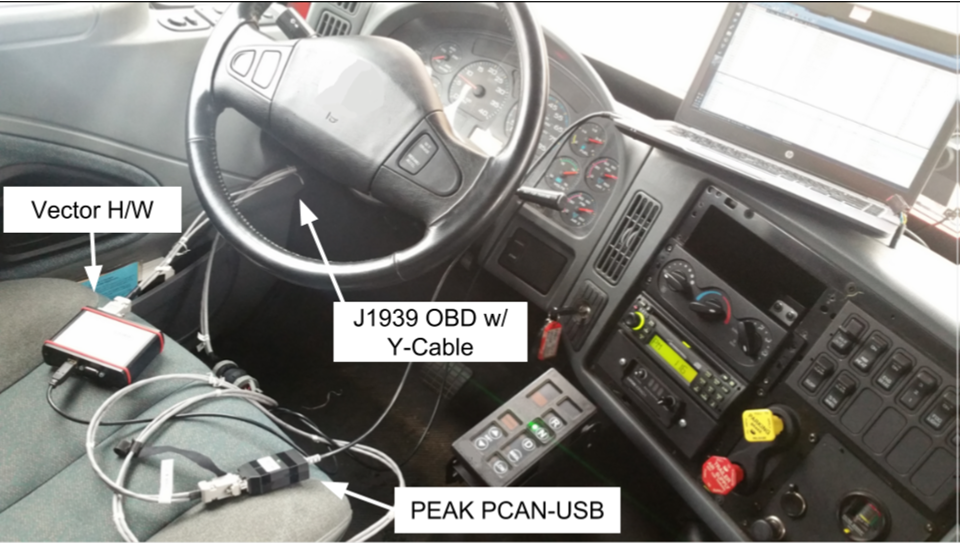}
        \caption{Example setup for experimentation in case \\ study 5: heavy-duty vehicles \cite{198487}.}
        \label{fig:truck}
\end{figure}

The compromised CAN-based standard \textit{J1939}, used across a wide variety of heavy-duty vehicles, supports vehicle telematics, e.g., diagnostics, maintenance control, fuel reduction, compliance checks, etc.~\cite{csselectronics_2020}. Specifically, telematics enable vehicles to communicate with the outer world through different telecommunication technologies; thus, potentially exploited vulnerabilities could enable granting adversaries access the vehicle's network. Through the \textit{J1939s} OBD port and leveraging Vector CANoe, an industry-standard CAN analysis and simulation software tool, one can perform a variety of attacks such as packet eavesdropping and packet injection. 
By parsing and inspecting the obtained data (via packet eavesdropping), the Parameter Group Numbers (PGN) -- which are message identifiers that control various functions of the vehicle -- can be identified and manipulated. These eavesdropping and traffic analysis attacks can be classified as \textit{Communication}-based IoT attacks. Exploiting the same port and utilizing the already-captured packets using PEAK USB-PCAN (an alternative packet analysis utility to Vector CANoe), packet injections were successfully performed on both the school-bus and the truck, also \textit{Communication}-based attacks. However, since a physical connection to the OBD port is required for this type of attacks, they fall under the category of \textit{Device} attacks. {Typical consequences of the CAN being compromised, can result in sudden increases of the vehicle's engine RPMs, unexpected brake engagement, and indication changes including oil pressure, service brake pressure, and other gauges.} Fig. \ref{fig:truck} illustrates the experimental compromise setup of the aforementioned CAN protocol attack. 

The security hazard of these attacks lies in the fact that attackers are not only able to override the driver's input, but also remove the user's ability to regulate the vehicle torque. 
Furthermore, attackers could disable the truck's engine breaking system when it is running below $30~mph$. Both attack scenarios can lead to disastrous results threatening the safety of the passengers as well as their surroundings. Another attack case with catastrophic compromise results is the malicious modification of the brake service pressure indicator, since without such display signal drivers do not have any other means to gauge the brake air pressure and avoid emergency all-wheel lock.
The authors of \cite{198487} do not suggest any specific mitigations for the described attacks. {Their aim was to raise awareness and express their concerns regarding the various wireless technologies being ported to the heavy-duty vehicle industry, and highlight that vehicle security should not be overlooked. Although, there are various studies that emphasize on the prevention and mitigation of attacks targeting vehicles, using state estimation and control algorithms that can potential encounter disastrous results \cite{8862884}}.

\subsubsection{Case Study 6 -- Healthcare Medical Devices} \label{ch:healthcare}

{The healthcare infrastructure covers a vast part of the commercial IoT pillar. Attacks targeting medical IoT devices can range from confidentiality-type compromises (e.g.,disclosure of user medical records) to healthcare system-wide events which can harm patients' well-being. During the COVID-19 crisis, the U.S Department of Health and Human Services~(HHS) reported a significant increase of 50\% more cyberattack incidents targeting the healthcare industry.} One contributing factor for this sheer cyberattack increase could be the fact that more than \%80 of the $1.2$ million IoT devices located in U.S healthcare organizations were operating using outdated OS which do not support security updates anymore~\cite{gregory_2020}.

In September 2020, a hospital in Germany sustained a ransomware attack and as a result was unable to admit a patient due to system access unavailability. 
Unfortunately, a patient in critical condition lost her life since she had to be transferred to another hospital almost $32km$ away. From a medical perspective, it can be argued that the distressful situation -- caused by the ransomware attack -- could have potentially contributed to the death of the patient. However, this speculation cannot serve as sufficient proof to legally convict the attackers for homicide~\cite{ralston_2020}. The incident was initiated due to a flaw in the VPN network; hospitals are known for owning outdated vulnerable systems \cite{erven_collao_2015}.  
Furthermore, due to the integration of new devices and the expansion of the healthcare infrastructure with medical IoT devices, a rise in attacks has been observed recently~\cite{hipaa_2019}. The situation is further exacerbated since 
the majority of such medical IoT devices are typically 
outdated and lack potent cybersecurity features, paving the way for adversaries attempting to penetrate into these mission-critical healthcare systems~\cite{newcomb_2020}. 

Notably, in 2015, researchers developed a search engine, named Shodan, for finding exposed devices connected to the Internet, for example, exposed medical IoT devices used in the healthcare system \cite{erven_collao_2015}. 
As a result,  
Internet-connected medical systems were exposed on a public domain, such as anesthesia systems, cardiology systems, infusion systems, magnetic resonance imaging~(MRI) machines, picture archiving and communication systems~(PACS), nuclear medicine and pacemakers. The majority of these systems were  inadequately protected against critical vulnerabilities like default credentials, emergency account login, Telnet-root access, FTP - Admin, SSL key password manager, etc. In essence, the systems were exposed to \textit{Communication} and \textit{Infrastructure} -type of attacks. The identified vulnerabilities of this incident were initially discovered between 2001 and 2013 and are included in the Common Vulnerabilities and Exposure (CVE) database, underlining the lack of cybersecurity considerations in healthcare organizations.  
With detailed information about the existing healthcare system and its exact location and functionality within the hospital network, attackers can exploit them to execute physical attacks and mount \textit{Device}-type of attacks. Using employees data (e.g., credentials), \textit{Service}-type of attacks like phishing and spearphising bolster high probability of success. Another crucial threat however, was that the network was unprotected against the remote code execution \textit{MS08-067} vulnerability~\cite{microsoft_2008} of Windows XP operating systems while also being connected to the back-end of the infrastructure. {In an effort to demonstrate the lack of cybersecurity countermeasures and lure adversaries, security researchers utilized honeypots to mimic the system as a target for adversaries.} Honeypot machines are mechanisms that replicate the behavior of deployed vulnerable devices while also providing history and traffic logging capabilities.

\begin{figure}[t]
    \centering
        \includegraphics[width=0.90\linewidth,trim={0cm 0cm 0cm 0cm}]{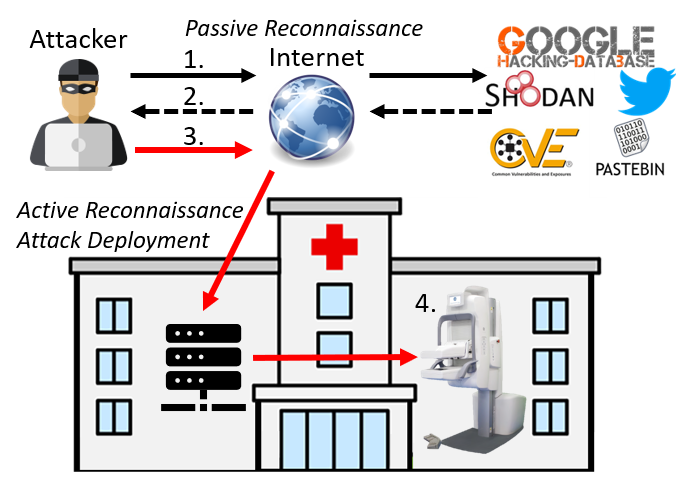}
        \caption{Healthcare attack procedure of case study 6.}
        \label{fig:healthcare}
\end{figure}

Using the exposed medical device characteristics (e.g., models, connections, IPs, etc.),  
the honeypots were installed mimicking the healthcare organization architecture. Any vulnerability present in the original medical IoT device was replicated in the corresponding honeypot machine creating a digital twin of the actual healthcare network. 
Eventually, the ``honeypot devices" were discovered by various Internet search engines and vulnerability databases. Six months later after the honeypot deployment, hackers discovered these counterfeit devices, exploited most of their vulnerabilities, gained access to various emulated medical devices and even left malware on the honeypots. Reviewing the honeypot findings, an attacker could have followed a similar approach and compromise the vulnerable internet connected medical IoT devices. Fig.  \ref{fig:healthcare} illustrates how 
an adversary could attain access to the emulated healthcare system environment. Specifically, the attack scenario could involve the following steps, \textit{(1)} perform passive reconnaissance using various tools, and \textit{(2)} get information about the Internet-connected devices. Then,  \textit{(3)} get important information regarding the devices with active reconnaissance, i.e., port scanning, a \textit{Communication} attack. Using the recorded device exploits, \textit{Service}-type attacks could be performed like brute force or dictionary attacks, achieving remote login access to these device via SSH or the device's web interface. Following, any other vulnerability existing in the simulated network could be exploited (FTP admin access, telnet root access, service login CVE-2009-5143). The command and control server could send commands to other systems threatening the \textit{Infrastructure}, e.g., install malicious software or firmware to the medical \textit{Devices}. At this point, the attacker could be assumed to have limitless capabilities due to access to the whole network system, the medical IoT devices \textit{(4)}, and their corresponding data. 
Furthermore, an attacker could also block legitimate user access to the healthcare system, potentially leading to similar events to the ransomware incident in September of 2020. 

It is important to note that, during the period that the honeypots were active, $55,416$ successful logins were established, $24$ successful exploitations were performed (the majority of them being the MS08-067), $299$ malware samples were dropped in total, and $8$ HoneyCreds logins were registered. {The report in \cite{erven_collao_2015} suggests that all healthcare organisations should check their Internet-connected IoT devices for default credentials, report identified issues to manufacturers requesting remediating actions, and incorporate cybersecurity practices in their existing systems. Most healthcare centers have machines that run outdated operating systems (e.g., Windows XP, Windows 2000, Windows 7) which lack security updates and therefore must be upgraded or replaced. The cybersecurity dimension of medical infrastructure, including commercial medical IoT devices, should be a core principle when designing and maintaining healthcare systems to evade incidents that violate data integrity and harm human safety.}

\subsection{Incidents Targeting Industrial IoT Devices}

IIoT systems combine industrial control systems~(ICS), IoT, and IT/OT technologies.  As a result, IIoT systems can harness advanced networking capabilities, allowing them to operate efficiently in a distributed fashion while also being monitored and controlled from remote and dispersed locations. 
However, these connected features can also arise as the ``weak links''. {Thus, if such connected IIoT systems are not properly secured, adversaries can exploit them leveraging remote connections through the Internet or by penetrating industrial networks \cite{gurtov2016secure}.} Attacks can also be performed by compromising industrial personnel accounts and misusing their privileges~\cite{Haber2020}. Such incidents have been reported in many industrial facilities, where IIoT systems can either be targeted directly by attackers (e.g., IIoT device attacks), or they can be compromised as a result of the interconnected nature of IIoT and IT systems (e.g., attacks propagating from the IT infrastructure to IIoT devices). In this part, we discuss three IoT attacks derived from the industrial domain, which if not properly managed, could have caused disastrous consequences and even cost human lives. {Specifically, we present attack incidents of ransomware, unauthorized access to water treatment facilities, and modify control logic type of attacks in petrochemical plants.} 

\subsubsection{Case Study 7 - Ransomware -- WannaCry} \label{ch:ransomware}

In May 2017, the ransomware \emph{WannaCrypt}, also known as \emph{WannaCry}, infected around $230,000$ computing systems globally. WannaCry encrypted the files of the infected machines requesting a $\$300$  ransom 
in order to provide users with access to the encrypted documents. If the ransom was not paid withing a certain time frame, the price was first raised to  $\$600$, and  if the user still refrained to cooperate, WannaCry would proceed and delete the data permanently after three days. 
Among the first victims of WannaCry were a Spanish mobile company as well as thousands of National Health Service~(NHS) hospitals and surgeries across the United Kingdom. The ransomware spread beyond the European borders infecting  numerous computer systems globally~\cite{kaspersky_2020}.

Pharmaceutical manufacturing companies like Merck, automotive corporations like Nissan and Renault, food manufacturers like Modelez, and many other industrial facilities were infected with WannaCry rendering their computer systems unresponsive and compromising their facilities~\cite{shohet_2019}. 
In 2018, a variant of WannaCry targeted TSMC's Integrated Circuit~(IC) fabrication facilities. TSMC  supplies IC chips to many technology companies such as Apple, Qualcomm, Nvidia, and AMD. 

The ransomware caused operational downtime that directly affected TSMC's production lines, resulting in approximately $\$170$ million cost  \cite{kirk_ross_2018}. Although industrial facilities were not the main target of WannaCry attacks, industrial systems  were affected due to their dependency on information technology~(IT) systems, and specifically unpatched Windows machines. Furthermore, WannaCry could spread through  enterprise networks affecting any connected device, regardless of being part of the IT or operational technology (OT) infrastructure \cite{shohet_2019}. The incorporation of IIoT within OT/IT infrastructures, despite enabling industrial systems operators to monitor and control  facilities operations,  also expands their vulnerability surface to malware with disastrous consequences, as demonstrated in the TSMC case. {The situation is further exacerbated since many of IIoT devices are expanding their capabilities incorporating OS to achieve multitasking, improve security, and enhance communication due to their lack of inter-operability~\cite{8320780}, rendering them prominent targets for malware attacks with disastrous consequences.}

The life-cycle of  ransomware consists of the following stages: \textit{ (i)}~deployment, \textit{(ii)} installation, \textit{(iii)} destruction, and the  \textit{(iv)}  command-and-control~(C\&C) stage~\cite{7745494}. WannaCry leverages the MS17-010 vulnerability of  Windows OS during its deployment phase \cite{microsoft_2017}, and uses the ``Eternalblue'' exploit to install the ``DoublePulsar'' backdoor implant tool for the malicious code injection and execution~\cite{bhat_2017}. MS17-010 allows remote code execution in vulnerable Windows machines if an attacker sends a specific message to a Microsoft Server Message Block 1.0 (SMBv1). SMB is a client-server Application/Presentation layer protocol which enables users request access to resources on a server (i.e sharing files, opening or editing files, using printers and ports)~\cite{smb}. Then,  WannaCry injects the binary file \texttt{launcher.dll} through the exploit and the backdoor. WannaCry then exploits the SMB  driver \texttt{srv2.sys} to attain access to the compromised devices and send the malicious payload. The \texttt{launcher.dll} file, which is only being  executed in memory leaving no traces (on the disk), serves as the loader for the executable,  \texttt{mssecsvc.exe} file. Thereafter, the  \texttt{launcher.dll} runs the executable as a regular system process, and the second phase of ransomware begins. 

During the installation phase, the target system is analyzed by the malware to determine if it is an actual computer or a virtual machine, and then a \emph{hardcoded} domain name is queried. 
Notably, this specific domain name served as a kill-switch allowing the attacker to remotely execute or terminate the malware. If the domain name was responding, this would signal WannaCry to terminate. However, if the domain name was not responsive, the second part of the installation phase continued planting the \texttt{tasksche.exe}. This executable file manages the resource loading for the malware, and the encryption environment establishment. While the \texttt{mssecsvc.exe} manages the \texttt{mssecsvc2.0} service for either the ``dropper''~phase which is the creation of the \texttt{taskche.exe} file or the infection process. The infection process involves the exploitation of the SMB protocol; by broadcasting packages to the connected network is able to identify vulnerable devices (through their open port 445). 
If such devices exist in the network, WannaCry checks whether the machine includes a Doublepulsar backdoor for payload deployment, and if not, the exploitation of the Eternalblue commenced to plant the backdoor. 

After the setup phase, the \texttt{launcher.dll} along with other auxiliary files are sent to infect the new victim. At the \texttt{taskche.exe} file path, a new directory is created where the essential files for the attack are extracted. These files contain, among others, the instructions for the decryption of the user files, the target address, and the routing (via onion) information for the ransom exchange between the attacker and the infected user. {The purpose of onion routing is because such protocols use the concepts of proxy redirection and layered mixed-key cryptography to hide the routing requests from the participants of the network except the originator, the infected machine, of the request.} In other words, it  provides anonymity for  attackers   and protects them from traffic analysis that could lead to their discovery \cite{Hooks06onionrouting}.

\begin{figure}[t]
    \centering
        \includegraphics[width=9.5 cm,trim={0cm 0cm 0cm 0cm}]{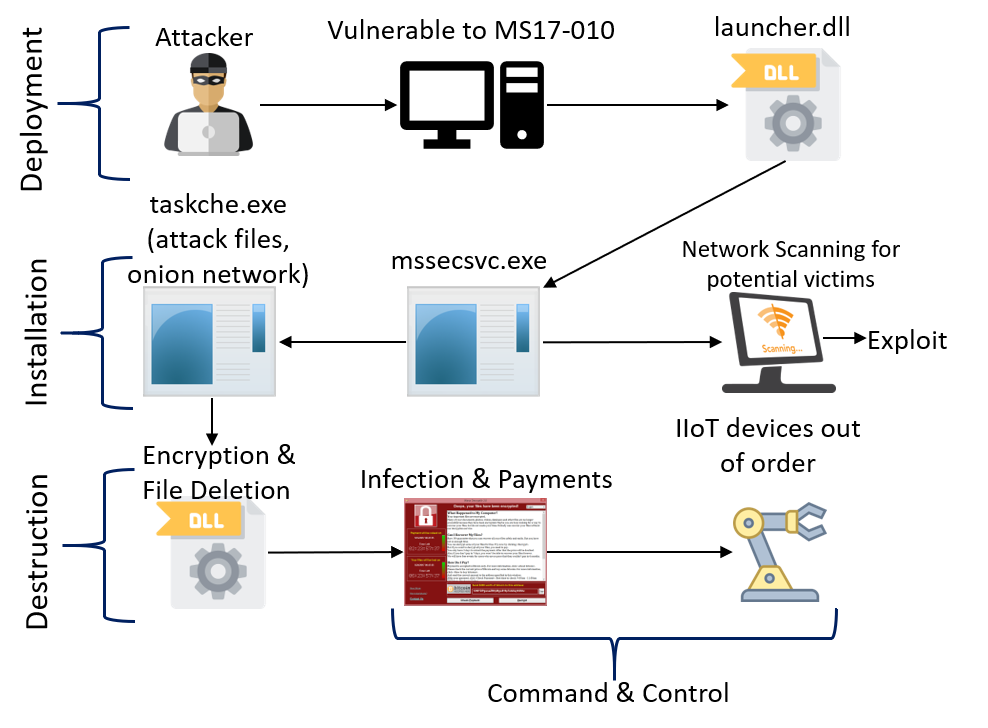}
        \caption{WannaCry attack flow in IIoT facilities.}
        \label{fig:WannaCry}
\end{figure}

At the destruction phase, WannaCry uses cryptographic algorithms  for the encryption of the victims' files.  Examples of such algorithms include RSA-2048 
and AES-128. 
After the data encryption, the device's unencrypted data are deleted, following a \emph{data wiping} procedure which overwrites data with random numbers, rendering any data restoration attempt futile. 
At the final phase, the \texttt{@WanaDecryptor@.exe} user interface gets deployed and executes a TOR service for the C\&C stage. The onion server supports the communication between users and the attacker, sending relevant information to the adversaries~(e.g. user name, host name, data of the infected system), and displaying the "infamous red-window" with instructions for the victim requesting to pay the ransom in order to recover the files~\cite{8323679}. The  user-friendly malware interface is made available in multiple languages and includes a tutorial demonstration permitting the decryption of up to $10$ user files at no cost. 
WannaCry's attack path is depicted in Fig.~\ref{fig:WannaCry}. The attack starts by exploiting an OS  vulnerability~(MS17-010), i.e., a \textit{Service}-type attack. Then, WannaCry uses the SMB protocol in a \textit{Communication}-type of attack to broadcast and find vulnerable devices with a planted Doublepulsar backdoor,  and eventually encrypts all the files and damages the whole \textit{Infrastructure}. By extension IIoT devices that were connected to the infrastructure got out of order due to the ransomware halting the whole production. Evidently, being able to compromise almost every aspect of a system in organisations~(i.e. Merck, Nissan, NHS, TSMC), it categorizes malware as one of the most threatening attacks. Malware attacks could not only impact the infected devices, but they could also propagate into the network and stealthily expand to neighbouring devices and systems and  cause multi-layered and system-wide compromises.
 

A researcher, Marcus Hutchins, while investigating the malware, accidentally activated its kill-switch. After identifying the domain to which WannaCry was registered, the researcher was able to deactivate it and disrupt its spread. Security analysts believe that the kill-switch was created to prevent malware probing in virtual sandbox environments~\cite{winckles_2018}. Therefore, the  modification of the malware code to utilize different domain kill-switches, enabled the generation of multiple variants of the malware  such as the one that halted TSCMs production line~\cite{shohet_2019}.

{WannaCry spread globally through phishing campaigns \cite{kaspersky_2020}, causing economic losses estimated around \$4 billion by compromising hospitals, firms, and big manufacturing and industrial facilities. To protect computer systems and underlined IT/OT infrastructure from such attacks, OS and software running on these machines should be updated regularly.} Furthermore, untrusted sources~(e.g., email attachments, websites, USB devices, etc.) should receive proper handling and from users for downloading, installing, or executing any software, or piece of code on the organisations' machines. Users should keep updated their security software~(e.g., antivirus)~\cite{kaspersky_2020} which can handle malware and backup their data, deploy firewall Intrusion Detection and Prevention Systems~(IDS,IPS), backup their data, increase the awareness in firms for possible scamming emails, deploy proper network segmentation which can prevent such spreading and monitor any malicious behavior in the network~\cite{trendmicro_2017}.

\subsubsection{Case Study 8 -- Water Treatment Facilities -- Kemuri} \label{ch:externalRemote}

A considerable increase of cyberattacks targeting water treatment and distribution critical infrastructure has been noticed during 2020.  
Three attacks have been reported targeting water treatment facilities located at Israel . The first attack aimed to modify the level of chemicals used to process tap water, while the other two targeted agricultural and city water pumps, respectively. Although minor damage was caused from the aforementioned cyberattacks, Israel's government mandated  water treatment facilities to enforce strict cybersecurity measures . Israel’s national cyber chief stated ``we can see something like this aiming to cause damage to real life and not to IT or data ''~\cite{paganini_2020}. 
Water and Wastewater Sector~(WWS) is considered as one of the most targeted lifeline infrastructure~\cite{dhsCIS}; thus, WWS is treated as a matter of national security. This conjecture is validated by the numerous cyberattack incidents targeting WWC which have been reported in the past years all over the world. Namely, fifteen such incidents are reported in~\cite{Hassanzadeh_2020}, including crypto-jacking, ransomware, backdoor deployment, and physical attacks from adversaries with diverse objectives, capabilities, and motives.

{One of the many incidents against the WWS, occurred in 2016 at an undisclosed water utility, commonly referred to using the pseudonym Kemuri Water Company~(KWC)~\cite{osti_1505628}. The KWC, after hiring a security firm to perform proactive cybersecurity assessment of their water supply and metering systems, was informed that various vulnerabilities existed in their systems. 
After thorough investigations, the security company found IP addresses of state-sponsored hacktivists in the facility's traffic reports and possible unauthorized access to the KWC subsystems along with a series of \emph{unexplained valve manipulation patterns}. After the forensics examination, further evidence was discovered in the KWC infrastructure indicating exfiltrations of $2.5$ million unique records as well as manipulation of the chemical flow rates~\cite{Hassanzadeh_2020}, similar to the incident in April 2020 at a water treatment facility~{\cite{paganini_2020}}. Specifically, the water district's endpoint OT systems were outdated running old OSs.} {For instance,  many critical IT/OT operations were operating on a single AS400, an IBM application system built in 1988 designed for small and intermediate-sized firms~\cite{ibm}.} The AS400 system was provisioning the following functionalities, \textit{(i)} operating as the SCADA platform, \textit{(ii)} a router for the various KMC connected networks, \textit{(iii)} controlling hundreds of PLC devices for the water valve and flow control applications, \textit{(iv)} hosting the personally identifiable information~(PII) and billing information of the water utility customers,  and \textit{(v)} storing the firms financial statements. In addition, the AS400 system was exposed to the Internet. 
The internal IP address and the admin credentials for the payment application webserver were stored in a plaintext \texttt{.ini} file, enabling access to adversaries. 
Furthermore, SQL injection vulnerabilities were discovered in the KWC payment portal~\cite{Vericlave}, in which multiple factor authentication was not required~\cite{verizonDataBreach}. 

\begin{figure}[t]
    \centering
        \includegraphics[width=8 cm,trim={0cm 0cm 0cm 0cm}]{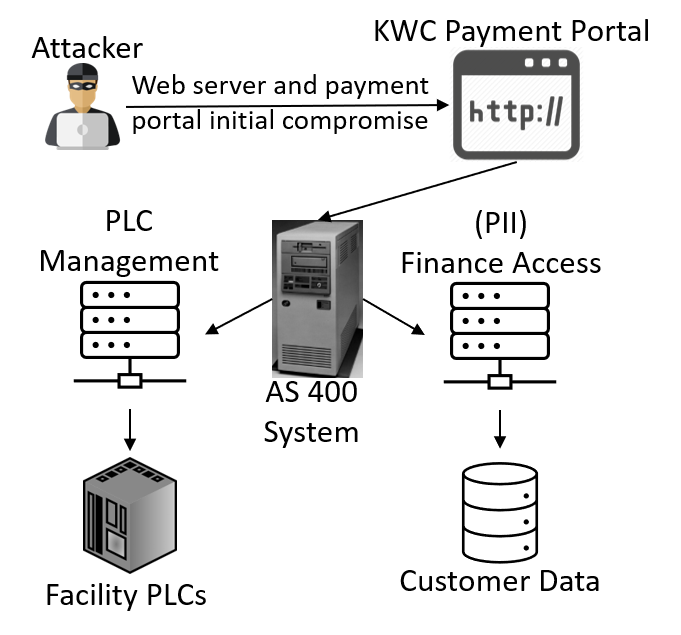}
        \caption{Adversarial path of the Kemuri attack incident.}
        \label{fig:kemuri}
\end{figure}

The attack path for the KWC incident is illustrated in Fig.~\ref{fig:kemuri}. 
The attackers exploited the limited security of the remote user login application, breaching the PII system and getting access to the client payment information~\cite{verizonDataBreach}. In particular, the authors in~\cite{Vericlave}, report that the initial compromise was indeed performed through the payment portal leveraging SQL injection and phishing attacks, which correspond to \textit{Service}-type attacks according to our proposed attack taxonomy. Next, the attackers leveraging the information within the plaintext \texttt{.ini} file of admin credentials for the web server, logged into the AS400 system. Granted access to the AS400, the attackers were able to seize about  $2.5$ million records from the PII database.  
Also, adversaries through the AS400 could impair the PLC management routines, threatening the overarching KWC \textit{Infrastructure} including PLCs, SCADA control signals managing water flow valves, chemical mixtures ratios, etc. As a result, the attackers could have tampered with the amount of chemicals intended for the water supply and handicapped the KMC water treatment and production capabilities resulting in operational downtime along with significant delays for the water reserve replenishment. Such PLC malicious modifications are considered \textit{Device}-type of attacks since they can directly affect the limits, set-points, and control strategies of actuators coordinating the industrial processes.

The identification of remote connection vulnerabilities from the investigation was addressed by terminating the account management front-end while any outbound connectivity from the AS400 was blocked as well. KWC was advised to replace legacy systems in order to conform to state-of-the-art security practises and standards, and patch the ones that can still support security updates. AS400 was identified to be the \emph{single-point-of-failure} as most processes were centralized around it. A segregated network architecture of distributed nature could alleviate such issue. 
The user authentication mechanisms should be strengthened  along with secure storage practises for user credentials. 
Additionally, intrusion detection and prevention schemes leveraging the data generated from IoT devices inside the facility's network could be deployed, as an additional layer of defense,  to preemptively probe for anomalous patterns and mitigate future breaches~\cite{Singh2020}.

{Critical infrastructure compromises such as the described incident at the water facility of KWC, could not only impact the utility operation on the financial sector, but also jeopardize people safety and health, e.g., if customers are supplied with contaminated water.}
In the KWC case, attackers penetrated through the IT infrastructure delivering the attack payload on the OT endpoints and tampering with the facility's management and water quality. The KWC attack incident demonstrates that  industrial facilities, and especially critical infrastructure, must enforce security strategies and standards while accounting for their cyber-physical nature and cross-layer attack propagation via IIoT components.

\subsubsection{Case Study 9 -- Modify Control Logic - Triton/TRISIS/HatMan} \label{ch:modifyControl}

Similar to the malicious code injection and  code modification attacks encountered in IoT systems, attacks in IIoT environments often target PLCs and are referred to as control logic modification attacks~\cite{themitrecorporation_2020}. In IIoT devices, the programmable code is typically written in ladder logic, functional block diagram~(FBD), or other PLC programming languages, instead of high-level languages, e.g., C/C++,  used in other embedded IoT devices. Another notable distinction of IIoT devices is that they coordinate physical processes, often in real-time, by controlling actuators on industrial equipment based on sensors input. The TRITON attack launched in 2017 targeted a petrochemical plant and 
belongs to this type of control logic modification attacks. The severity of TRITON is depicted by the fact that is often referred as ``the world’s most murderous malware''~\cite{MIT_triconex}. 

\begin{figure}[t]
    \centering
        \includegraphics[width=8.7 cm,trim={0cm 0.3cm 0cm 0cm}]{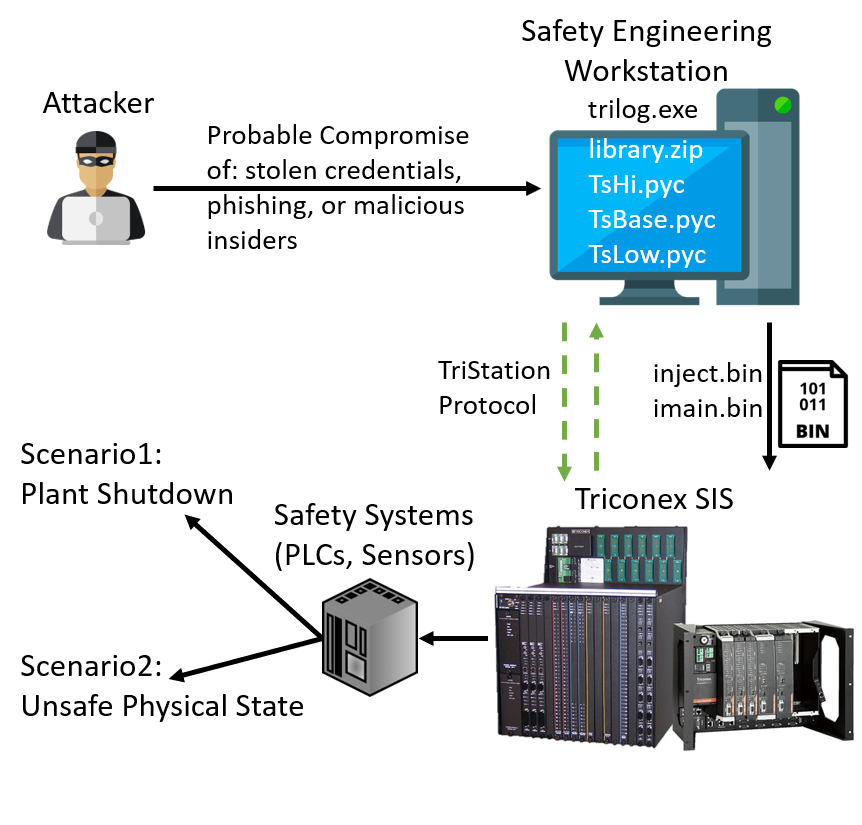}
        \caption{Triton attack flow scenarios.}
        \label{fig:tritonFLow}
\end{figure}

In the TRITON incident, attackers targeted the Safety Instrumented System~(SIS) workstation which is responsible for maintaining the nominal operation of the ICS, issue warnings, or even stop the process if safety limits are violated. After the attackers gained remote access to the SIS workstation, the TRITON malware was deployed to reprogram the connected IIoT controllers~\cite{johnson_2017}. Additionally, adversaries modified the firmware of critical PLCs, manipulated legitimate processes, and installed a Remote Access Trojan~(RAT) enabling them to access frequently the ICS infrastructure while hiding their existence. The potential attack path that adversaries exploited to mount their attack is demonstrated in Fig. \ref{fig:tritonFLow}. Fortunately, the attack failed without causing any physical harm to the personnel working on-site, or severely damaging the facility. The attack accidentally triggered an automatic shutdown causing a minor disruption to the plant, notifying the system operators to investigate the incident. Although in this occurrence the attackers ``helped'' identify the attack before more serious events jeopardizing human lives occurred, the stealthiness and sophistication of the attack justifiably alarmed the ICS industry~\cite{virsec}.

TRITON, TRISIS or HatMan\footnote{Named TRITON by FireEye~\cite{johnson_2017}, TRISIS by Dragos Inc.~\cite{dragos_2017} because it was targeting the SIS engineering workstation and HatMan by the DHS~\cite{dhsics}} attack targets Schneider Electric Triconex SIS controllers communicating using the TriStation protocol. Although the entry point of TRISIS is still not determined, the authors of~\cite{virsec} believe that the attackers attackers exploited stolen credentials via phishing campaigns or through malicious insiders in order to access and compromise the SIS engineering workstation. We can classify such type of attacks aimed at IIoT systems as \textit{Service} or \textit{Device} attacks. As a result, the attackers were granted access to the process control network in which supervisory human machine interfaces~(HMI) monitor the plant's distributed control systems  and engineering workstations~\cite{dragos_2017}. Then the SIS TriStation was infected with an \verb|x86| executable, named  \verb|trilog.exe|, which mirrored a TriLogger application and was the main actor of the attack. The counterfeit TriLogger was able to to record, playing back, and analyze or maliciously modify high-speed operating data from Triconex controllers~\cite{accenturesecurity}. The malware was programmed to bypass whitelist filtering and, by tweaking monitoring mechanisms, allowed the execution of specified software. TRISIS utilized a set of sub-modules stored in an archive (\verb|library.zip|) containing standard Python libraries as well as some modules (e.g, \verb|TsHi.pyc, TsBase.pyc, TsLow.pycm|) with capabilities to process, manipulate controls,  and directly communicate with PLCs. By leveraging the aforementioned scripts, the plant  \textit{Infrastructure} was stealthily compromised. For example, one of the scripts included in the \verb|trilog.exe| file search for vulnerable firmware on certain controllers. {Since the ICS systems work with triple redundancy, i.e., the three endpoints of the system should perform identically before any changes can be applied, a failure to program or update three PLCs concurrently  triggers an alarm.} This makes the vulnerable firmware script crucial since it has to maintain the triple redundancy principle at all times. 
          
{Moreover, the attackers, by reverse engineering the communication protocols used between the workstation and the controllers, were also able to mimic legitimate workstations and perform \textit{Communication}-based attacks. In the scenario which the PLCs were in programming mode, which is activated using a physical key allowing firmware updates, malicious commands and/or backdoors can be injected into the plant communication network. The attacked PLCs were meticulously programmed to hide any traces indicating a firmware update while the memory payload existed for a limited time.} Next, a RAT was installed on the controllers (\textit{Device}), granting attackers with ad-hoc access. Notably, utilizing \verb|inject.bin| and \verb|imain.bin| files, the \verb|trilog.exe| could construct a payload consisting of controller logic byte-codes and malicious functions able to bypass controller code-checking mechanisms. Considering that, privilege escalation on the controllers could be performed, granting adversaries with full access to the  memory of the Triconex platform. {Exploiting two zero-day vulnerabilities, i.e., ~CVE-2018-7522~\cite{nvd_2018} and~CVE-2018-8872~\cite{nvd_2018_2}, attackers gained elevated privileges on the controllers and executed arbitrary codes at the PLCs of the SIS system via the compromised TriStation.} 

If the TRITON attack was successful in 2017, one of the following two scenarios would likely occur; either an operational plant shutdown or the plant would operate in an unsafe state~\cite{dragos_2017}. 
The first scenario would cause a false-positive trigger with an unexpected shutdown while the system would not be in a ``dangerous'' state and eventually expose the attack after the analysis of the occurrence. The second scenario would allow the system to operate under unsafe conditions and concurrently continue to function properly endangering many cyber-physical components of the facility along with the employees of the facility, inhibiting situational awareness and rendering the deployed safety mechanisms purposeless. 
The plant shutdown would mainly cause financial losses during the offline time, and would require specific procedures in order to start the whole system without damaging any physical equipment. In the unsafe operational state scenario, besides the economic losses, the production line, the physical equipment, or even the safety of plant personnel could be jeopardized. The attackers of TRISIS while connected to the compromised system, wrote accidentally into a PLC memory location using a wrong format, leading two of the three controllers to operate abnormally. The triple redundancy was not full-filled and that caused an immediate safety system shutdown and by extension the discovery of the attack.

{Among the suggestions to avoid similar incidents include segregating safety system networks from process control and information networks by maintaining them isolated, and keeping the TriStation terminals locked into cabinets while allowing only connections to the safety network.} Proper physical controls should be in place for avoiding any unauthorized access to critical operations. For example, Triconex controllers could require a physical key in order to program them. {Also, the changes of the key states should be audited issuing alerts when the PLCs are set in programming mode.} Additional security suggestions include the use of unidirectional gateways, blocking bidirectional network connections for applications requiring information from the SIS. Furthermore, the implementation of access control and application whitelisting could be required for any user or service attempting to reach the SIS through the Internet. Any mobile data exchanged method~(i.e. USB flash drive, external Hard Drive) should be scanned prior connecting to the TriStation terminals, and devices that have been connected to other networks 
should undergo a proper digital sanitizing to detect any malicious software that they may have and could endanger the system.  Finally, monitoring of the ICS network traffic for any suspicious activity~(i.e abnormal behavior in the network, IP white listing) must be performed regularly~\cite{dragos_2017, johnson_2017}.

{\em Attack Mapping Discussion:} 
{ The discussed attack use cases targeting IoT architectures in consumer, commercial, and industrial environments are summarized in Table \ref{tab:table2}. It is important to reiterate that most of the described attacks can be attributed to more than one attack taxonomy category. {Hence, in Table \ref{tab:table2}, we demonstrate not only the classification of each IoT case study to the corresponding category, but also the attack pertinence (if any) to the rest of the attack categories. In more detail, the circles underneath each attack category indicate the relevance of each case study to a specific category. A white circle indicates no relevance at all, while a totally black circle designates the complete alignment of the case study to this specific category.} }

\begin{table*}[t]
\centering
\caption{{Mapping of attack case studies to the proposed taxonomy.}}
\label{tab:table2}

\begin{tabular}{||l|l|c|c|c|c||}

\hline \hline
 & {\textit{\textbf{Case Study (A/A)}}}  & \textbf{Device} & \textbf{Infrastructure} & \textbf{Communication} & \textbf{Service} \\ \hline
\multirow{3}{*}{\textbf{Consumer IoT}}   & 1: Voice Assistant               & \pie{0}  & \pie{90} & \pie{90}  & \pie{180}\\
                                         & 2: Baby Monitoring Cameras       & \pie{90} & \pie{90} & \pie{180} & \pie{90}   \\
                                         & 3: The Mirai Botnet              & \pie{90} & \pie{180}& \pie{90}  & \pie{90}  \\ \hline
\multirow{3}{*}{\textbf{Commercial IoT}} & 4: Aircraft Avionics             & \pie{0}  & \pie{90} & \pie{270} & \pie{90} \\
                                         & 5: Heave-Duty Vehicles           & \pie{180}& \pie{0}  & \pie{180} & \pie{0}   \\
                                         & 6: Healthcare Medical Devices    & \pie{180}& \pie{90} & \pie{180} & \pie{90} \\ \hline
\multirow{3}{*}{\textbf{Industrial IoT}} & 7: WannaCry Ransomware           & \pie{0}  & \pie{90} & \pie{90}  & \pie{180}\\
                                         & 8: Kemuri Water Facility         & \pie{90} & \pie{90} & \pie{0}   & \pie{180} \\
                                         & 9: The Trisis Malware            & \pie{90} & \pie{0}  & \pie{90}  & \pie{90}  \\ \hline \hline
\end{tabular}
\end{table*}

\section{Open Challenges}\label{s:challenges}

{The rapid penetration of IoT devices along with the diversity, heterogeneity, and the multitude of applications that IoT components span, distinguish them from traditional connected devices and networks. As a result, the security issues encountered in IoT interconnected networks, edge devices and low-cost IoT nodes diverge significantly from the ones in the IT security field. To to deal with existing IoT security challenges requires coordinated efforts from manufactures, governments, policy-enforcing agencies, and end-users.} In this section, we delineate the security challenges with particular emphasis on the case studies discussed in Section \ref{s:attackIncidents} and the corresponding vulnerabilities exploited in such scenarios. 

\begin{itemize}[leftmargin=*, wide=0pt]
    \item Security challenges in IoT ecosystems can be in part contributed to the plethora of vendors developing devices not considering potential vulnerabilities and their corresponding consequences if maliciously exploited. Additionally, negligent end-users who overlook security practises (e.g., do not patch their devices, use default credentials, etc.) undermine the security of IoT devices. For instance, in 2020, multiple vulnerabilities have been identified in 7 different TCP/IP open-source stacks (5 of which exist for almost 20 years now) \cite{newman_2020}. The presented case studies \ref{ch:baby_monitoring}, \ref{ch:mirai}, and \ref{ch:healthcare} clearly indicate that default credentials is a major factor jeopardizing the secure operation of IoT systems. To address this issue, government agencies have formed relevant policies, and IoT manufacturers have started deploying devices with randomized credentials. For example, the United Kingdom (UK) Minister of Digital announced recently that all pre-programmed passwords of IoT devices need be unique and should not return to their initial state of credentials with a factory reset, banning default passwords in IoT devices \cite{warman_2020}. Although such policies can help improve the security of newly manufactured IoT devices, already deployed systems could still be vulnerable to attacks initiated from default credentials. {Therefore, security awareness of end-users is crucial for compliance to such nondefault factory password policies.}
    
    \item The low cost of many IoT devices is a double-edged sword. On one hand, it drives their rapid penetration, but on the other hand, it should not be prioritized above the devices' security. {As presented in case study \ref{ch:mirai}, thousands of devices from all IoT pillars -- with the majority of them belonging to the consumer IoT domain -- were compromised.} Most of those IoT devices are inexpensive products which we use on a daily basis, such as printers, routers, IP cameras, etc., that lack fundamental security protections. Thus, IoT manufacturers should balance security performance and low cost and strive to produce secure devices before deploying them to the market. Future-proofing IoT devices by providing software updates to overcome the discovered vulnerabilities also proves challenging, since it introduces significant costs for IoT suppliers. For instance, many mobile phone suppliers discontinue issuing software support updates to their devices after 3-5 years. Evidently, issuing updates to low-cost IoT devices can become cost-prohibitive. In addition, depending on the application field of IoT platforms updating them can also be impractical since it might require specialized personnel with on-site or physical access.
    
    \item The computational power of IoT devices can also introduce security challenges. For example,  sophisticated encryption and authentication mechanisms might not be supported by resource-limited, low cost or even disposable IoT nodes.  Legacy protocols lacking encryption are often widely used in IIoT (e.g., Modbus). 
    In addition, industrial systems could use specific tailor-made protocols -- as it was in the case of TriStation in \ref{ch:modifyControl}, for which there is no public information detailing their structure. The \emph{security by obscurity} concept, relying on the secrecy of the protocol details, does not provide any formal security for the underlined protocol, and should be avoided especially in mission-critical deployments. 
    On the other hand, hardware security schemes  such as cryptographic processors, physical unclonable functions~{(PUFs)}, HMACs and random key generators arise as viable alternatives due to their low overhead and cost \cite{9155002, 9283536}. However, the main disadvantage of hardware security mechanisms is that they need to be incorporated during the IoT device design phases. Retrofitting them to legacy or deployed devices is often infeasible \cite{muyeen2017communication}.
    
    \item Even in the cases where IoT nodes are equipped with sufficient computational power and data processing units, security challenges still exist. The proliferation of smart IoT sensors is popularizing machine intelligence and learning-based methods in many critical CPS \cite{zografopoulos2021cyberphysical, konstantinou2015cyber}. The rise of learning-based schemes is accompanied by important security challenges: it creates an incentive among adversaries to exploit potential vulnerabilities of the algorithms. {The success of artificial intelligence and machine learning has been thwarted by adversarial attacks such as decision-time and data poisoning, which tend to introduce vulnerabilities into the learning process \cite{9087789, 9281719, iet-stg.2020.0015}.} We need to revisit the challenging problem of developing secure and robust learning-based algorithms utilized in IoT networks. {In addition, it is of paramount importance to start providing a thorough security assessment of existing learning-based techniques targeting the identification process of IoT node faults and the detection of malicious attacks. This could allow researchers to develop learning-based techniques by fusing domain-aware knowledge of the underlying IoT system nature into the learning model. The security-enhanced and robust mechanisms -- in the presence of motivated and sophisticated adversaries -- should still be efficient for realistic implementation in IoT applications.} Towards improving effectiveness, federated and reinforcement learning schemes can utilized in order for the the training to be performed on distributed data residing on intelligent electronic devices of the IoT network \cite{9261465}.
    
    \item The amount of data aggregated by IoT devices requires moderation and secure handling. {IoT manufacturers could regulate the amount of data that their devices collect, and implement security and access control mechanisms to protect the confidentiality  and access to user information.} At the same time, users should be aware of the data collected by IoT devices (e.g., cookies, geolocation data, etc.),  allowing them to decide whether they want their activity to be monitored, proactively protecting their security and privacy. Identifying the minimum amount of data required to enhance user experience, while effectively securing them from malicious adversaries is another obstacle that future IoT devices will have to overcome. Extreme caution should be exercised when interacting and storing user medical data (as discussed in \ref{ch:healthcare}), since adversaries have targeted healthcare organizations in multiple occasions.
    
\end{itemize}

As the penetration of IoT devices in our everyday lives continues to increase, it becomes imperative to identify the associated technical, economic and regulatory challenges, and to develop impactful solutions to ensure compatibility with the existing technological advancements and a smooth transition to secure, reliable, and resilient future CPS.
{A multitude of aspects should be considered when facing the challenging task of securing IoT-related assets. Apart from the threats posed to the IoT application domain, to services delivery continuity, safety and, in general, management risk, having improved security awareness for the weakest part of an IoT ecosystem is critical. Thus, humans' psychological flaws should be regarded as an organic part of the IoT security equation. }
\section{Concluding Remarks} \label{s:conclusion}

{In this paper, we demonstrate an attack taxonomy architecture designed with real-world IoT attack incidents in mind. 
We divide our attack taxonomy into categories and map IoT attacks to their corresponding attack class (Table \ref{tab:table2}). Furthermore, we disclose the underlined security vulnerabilities of the investigated IoT attacks and propose potent countermeasures which -- if enforced -- can subvert such vulnerabilities from commencing to full-blown attacks. 
Additionally, we examine three different IoT domains, i.e., the consumer, commercial and industrial sectors, given their diverse operational objectives and constraints, e.g., asset security, real-time operation, device life-cycles, etc. 
For each sector, we dissect three real-world attack incidents delineating the vulnerabilities and attack paths that adversaries exploited to mount them and map them to our taxonomy. We provide mitigation strategies and security recommendations to overcome the discussed attacks, as well as potential future attacks targeting similar IoT devices, operating in similar ecosystems  while harboring similar vulnerabilities. {Our attack taxonomy enables the systematic investigation of attack clusters (i.e., device, infrastructure, communication, service) instead of specific attacks -- overcoming the requirement to singularly investigate every newly encountered attack -- thus, expediting security retribution.}}

{Although the impact of IoT attacks can range depending on the targeted IoT device and the operational environment, certain vulnerabilities exist in a variety of IoT systems, e.g., default credentials, lack of network segregation, etc., regardless of the application sector. These critical vulnerabilities, which span the IoT spectrum, require specific handling and precise remediation. IoT device vendors, protocols, and users alone cannot safeguard the security of IoT ecosystems, thus unified security policies need to be implemented to enhance the security of consumer, commercial and industrial systems. 
Following this policy-oriented approach, researchers have proposed frameworks which can enhance the security of commercial IoT systems \cite{ngwenya_ngoepe_2020}. 
{Furthermore, tech firms have also joined this effort to modernize the IIoT sector (due to its mission-critical objectives), hardening IIoT against cyberattacks, improving its resilience and safeguarding its operations \cite{IIoTFramework, konstantinou2021towards}.} Moving forward, synergies between the academia, the industry along with state and nation-wide policy enforcing mechanisms are necessary to improve the overall security posture of the constantly-growing IoT systems. The unexpected influx of connected devices due to the COVID-19 pandemic accentuated the need for consolidated security practices \cite{europeancommission_2020}. As a result, the European Union (EU) immediately issued security strategies and policies in order to regulate and enforce security measures. The architectural heterogeneity, distributed, and interconnected nature of IoT systems practically prohibits the development of a \emph{be-all and end-all} security solution. However, leveraging the accumulated experience and knowledge of security engineers and scientists from the academic community and the industry arises as a good approach to safeguard the IoT. {The consolidated effort of security policy makers and security researchers can help even the security battlefield by effectively dealing with existing vulnerabilities, and impeding the propagation of newly discovered zero-days before they develop into threats or attacks.}}

\ifCLASSOPTIONcaptionsoff
  \newpage
\fi

\balance
\bibliographystyle{IEEEtran}
\bibliography{biblio}

\begin{thebibliography}{100}
\bibitem{MIT_triconex}
``Triton is the world’s most murderous malware, and it’s spreading.''
\end{thebibliography}

\begin{IEEEbiography}[{\includegraphics[width=1.1in,height=1.25in,clip,keepaspectratio]{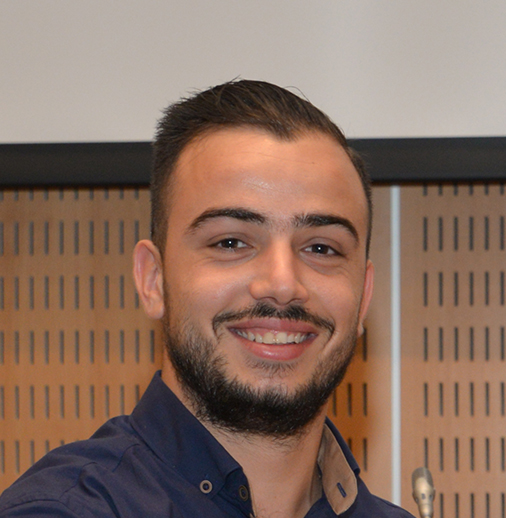}}]
{Christos Xenofontos} (Student Member, IEEE) received his BSc degree in Electrical Engineering from the department of Electrical and Computer Engineering at University of Cyprus, Nicosia, Cyprus in 2019. He is currently pursuing a MSc degree in Computer Science following the Computer Security study line at Technical University of Denmark, Kongens Lyngby, Denmark.
Additionally he is working part-time in the SQA team of Silicon Labs, Copenhagen in protocol testing for the Z-Wave protocol stack. His interests include Computer Security, with particular focus on IoT security and security in Cyber Physical Systems.
\end{IEEEbiography}

\begin{IEEEbiography}[{\includegraphics[width=1in,height=1.25in,clip,keepaspectratio]{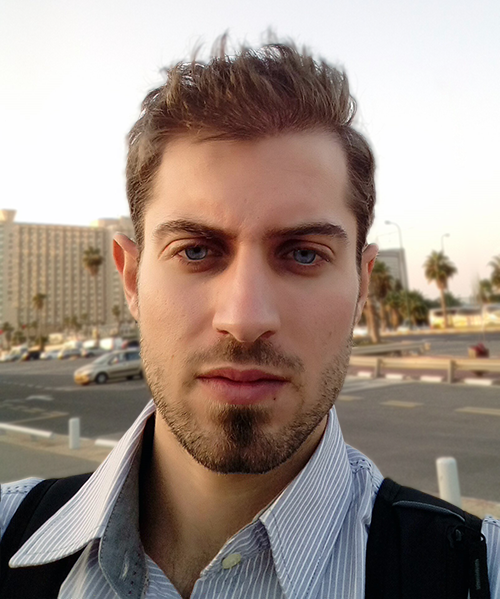}}]{Ioannis Zografopoulos}
received the B.Eng. and M.Eng. degrees in Computer, Communications, and Network Engineering, and the M.Sc. degree in Electrical and Computer Engineering from the University of Thessaly, Volos, Greece, in 2014 and 2015, respectively. He is currently pursuing the Ph.D. degree at the Computer, Electrical and Mathematical Sciences and Engineering Division (CEMSE) of King Abdullah University of Science and Technology (KAUST). His research interests include cyber-physical and communications security, with emphasis on IoT and embedded systems for industrial, distributed energy, and power grid applications. He is an IEEE and IEEE PES graduate student member and has served as a reviewer for IEEE Consumer Electronics Magazine, IEEE Computer Society Journal, as well as a sub-reviewer for ISVLSI and CPSIoTSec.
\end{IEEEbiography}

\begin{IEEEbiography}[{\includegraphics[width=1.1in,height=1.25in,clip,keepaspectratio]{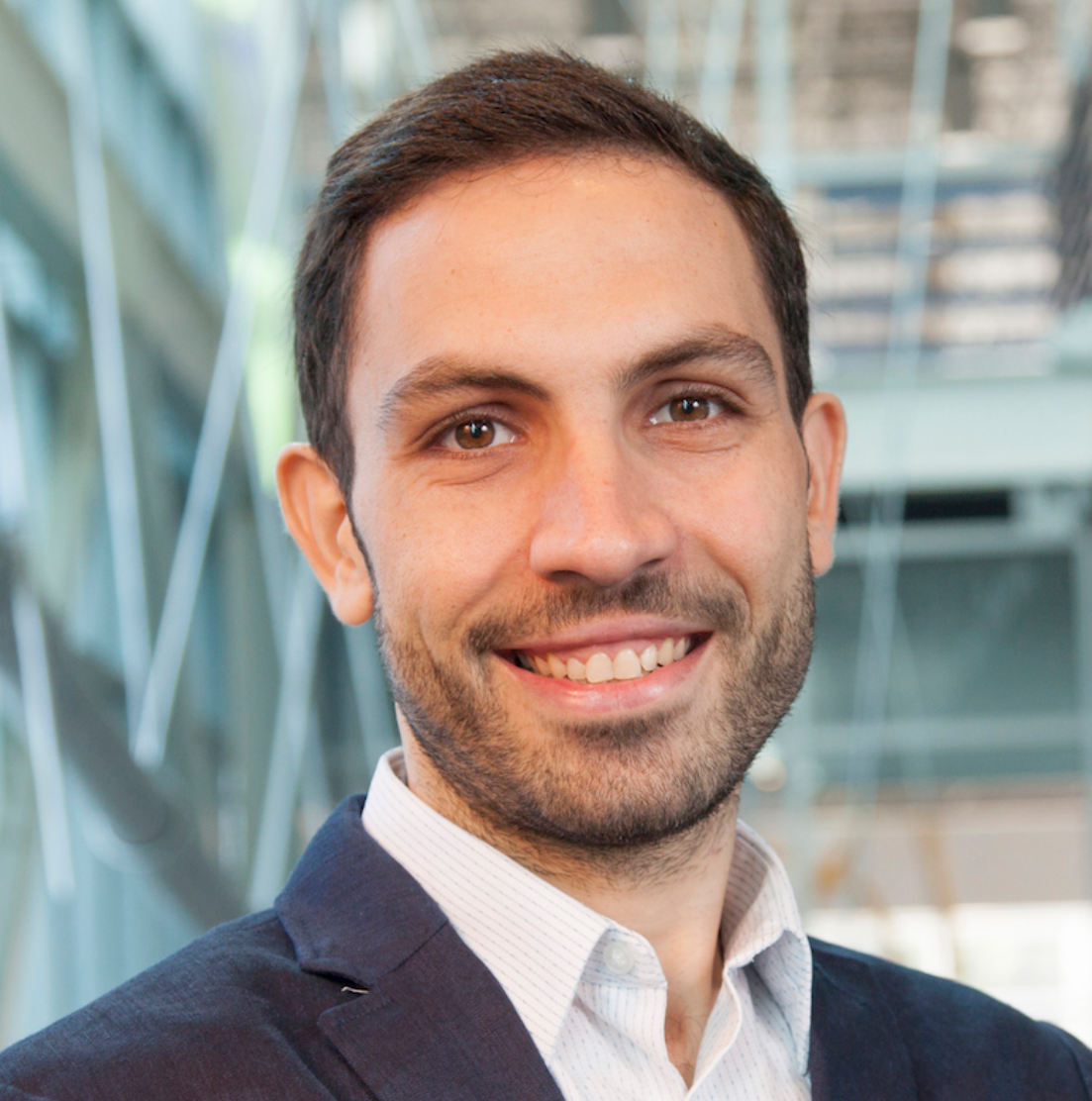}}]
{Charalambos Konstantinou} is an Assistant Professor at the Computer, Electrical and Mathematical Science and Engineering Division (CEMSE) of King Abdullah University of Science and Technology (KAUST), Thuwal, KSA. Before joining KAUST, he was an Assistant Professor with the Center for Advanced Power Systems (CAPS) at Florida State University (FSU), Tallahassee, FL. He received a Ph.D. in Electrical Engineering from New York University, NY, in 2018, and a Dipl.-Ing.-M.Eng. Degree in Electrical and Computer Engineering from National Technical University of Athens (NTUA), Greece, in 2012. His research interests include cyber-physical and embedded systems security with focus on power systems. He has authored multiple articles in the IEEE/ACM transactions and conference proceedings, and serves in the program committee of several international conferences. He is currently the Chair of the IEEE Task Force on Resilient and Secure Large-Scale Energy Internet Systems and the Secretary of the IEEE Task Force on Cyber-Physical Interdependence for Power System Operation and Control. He is the recipient of the 2020 Myron Zucker Student-Faculty Grant Award from IEEE Foundation, the Southeastern Center for Electrical Engineering Education (SCEEE) Young Faculty Development Award 2019, and the best paper award at the International Conference on Very Large Scale Integration (VLSI-SoC) 2018. He serves as the (Guest) Associate Editor for the International Journal of Electrical Power \& Energy Systems, IEEE Computer, and IEEE Consumers Electronics Magazine. He is a Senior Member of IEEE, a member of ACM, and an ACM Distinguished Speaker.
\end{IEEEbiography}

\begin{IEEEbiography}[{\includegraphics[width=1.1in,height=1.25in,clip,keepaspectratio]{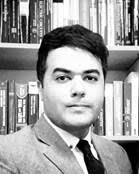}}]
{Alireza Jolfaei} received the Ph.D. degree in Applied Cryptography from Griffith University, Gold Coast, Australia. He is the Program Leader of Master of IT in Cyber Security at Macquarie University, Sydney, Australia. His main research interests are in Cyber and Cyber-Physical Systems Security. He has participated in several projects involving different aspects of Cyber Security. On these topics he has published over 60 papers appeared in journals, conference proceedings, and books. Before Macquarie University, he has been a faculty member with Federation University Australia and Temple University in Philadelphia, PA, USA. He has received multiple awards for Academic Excellence, University Contribution, and Inclusion and Diversity Support. He received the prestigious IEEE Australian council award for his research paper published in the IEEE Transactions on Information Forensics and Security. He served as the Chairman of the Computational Intelligence Society in the IEEE Victoria Section and also as the Chairman of Professional and Career Activities for the IEEE Queensland Section. He has served as the associate editor of IEEE journals and transactions, including the IEEE IoT Journal, IEEE Sensors Journal, IEEE Transactions on Industrial Informatics, IEEE Transactions on Industry Applications, IEEE Transactions on Intelligent Transportation Systems, and IEEE Transactions on Emerging Topics in Computational Intelligence. He has served as a program co-Chair, a track Chair, a session Chair, and a Technical Program Committee member, for major conferences in Cyber Security, including IEEE TrustCom. He was the General Chair of the 6th IEEE International Conference on Dependability in Sensor, Cloud, and Big Data Systems and Applications (DependSys 2020) in Fiji. He is a Distinguished Speaker of the Association for Computing Machinery (ACM) on the topic of Cyber Security and a Senior Member of the Institute of Electrical and Electronics Engineers (IEEE). 
\end{IEEEbiography}

\vspace{-0.3em}

\begin{IEEEbiography}[{\includegraphics[width=1.1in,height=1.25in,clip,keepaspectratio]{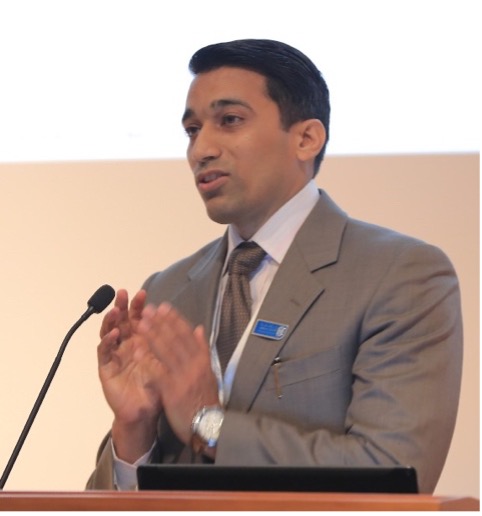}}]
{Muhammad Khurram Khan} is currently working as a Professor of Cybersecurity at the Center of Excellence in Information Assurance, King Saud University, Kingdom of Saudi Arabia. He is founder and CEO of the ‘Global Foundation for Cyber Studies and Research’ (http://www.gfcyber.org), an independent and non-partisan cybersecurity think-tank in Washington D.C, USA. He is the Editor-in-Chief of ‘Telecommunication Systems’ published by Springer-Nature with its recent impact factor of 1.73 (JCR 2020). He is on the editorial board of several journals including, IEEE Communications Surveys \& Tutorials, IEEE Communications Magazine, IEEE Internet of Things Journal, IEEE Transactions on Consumer Electronics, Journal of Network \& Computer Applications (Elsevier), IEEE Access, IEEE Consumer Electronics Magazine, PLOS ONE, and Electronic Commerce Research, etc. He has published more than 400 papers in the journals and conferences of international repute. In addition, he is an inventor of 10 US/PCT patents. He has edited 10 books/proceedings published by Springer-Verlag, Taylor \& Francis and IEEE. His research areas of interest are Cybersecurity, digital authentication, IoT security, biometrics, multimedia security, cloud computing security, cyber policy, and technological innovation management. He is a fellow of the IET (UK), a fellow of the BCS (UK), and a fellow of the FTRA (Korea). He is the Vice Chair of IEEE Communications Society Saudi Chapter. He is a distinguished Lecturer of the IEEE. His detailed profile can be visited at http://www.professorkhurram.com
\end{IEEEbiography}

\begin{IEEEbiography}[{\includegraphics[width=1.1in,height=1.25in,clip,keepaspectratio]{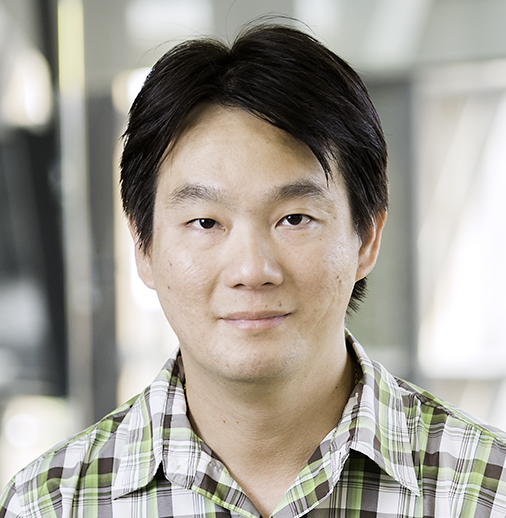}}]
{Kim-Kwang Raymond Choo} received the Ph.D. in Information Security in 2006 from Queensland University of Technology, Australia. He currently holds the Cloud Technology Endowed Professorship at The University of Texas at San Antonio (UTSA), and has a courtesy appointment at the University of South Australia. He is the founding chair of IEEE Technology and Engineering Management Society (TEMS)’s Technical Committee on Blockchain and Distributed Ledger Technologies, and serves as the Department Editor of IEEE Transactions on Engineering Management and the Associate Editor of IEEE Transactions on Dependable and Secure Computing, and IEEE Transactions on Big Data. He is an ACM Distinguished Speaker and IEEE Computer Society Distinguished Visitor (2021 - 2023), and included in Web of Science's Highly Cited Researcher in the field of Cross-Field - 2020. In 2015, he and his team won the Digital Forensics Research Challenge organized by Germany's University of Erlangen-Nuremberg. He is the recipient of the 2019 IEEE Technical Committee on Scalable Computing Award for Excellence in Scalable Computing (Middle Career Researcher), the 2018 UTSA College of Business Col. Jean Piccione and Lt. Col. Philip Piccione Endowed Research Award for Tenured Faculty, the Outstanding Associate Editor of 2018 for IEEE Access, the British Computer Society's 2019 Wilkes Award Runner-up, the 2014 Highly Commended Award by the Australia New Zealand Policing Advisory Agency, the Fulbright Scholarship in 2009, the 2008 Australia Day Achievement Medallion, and the British Computer Society's Wilkes Award in 2008. He has also received best paper awards from the IEEE Systems Journal in 2021, IEEE Consumer Electronics Magazine for 2020, EURASIP JWCN in 2019, IEEE TrustCom 2018, and ESORICS 2015; the Korea Information Processing Society's JIPS Outstanding Research Award (Most-cited Paper) for 2020 and Survey Paper Award (Gold) in 2019; the IEEE Blockchain 2019 Outstanding Paper Award; and Best Student Paper Awards from Inscrypt 2019 and ACISP 2005. His research has been funded by NASA, National Security Agency, National Science Foundation, U.S. Department of Defense, CPS Energy, LGS Innovations, MITRE, Texas National Security Network Excellence Fund, Australian Government National Drug Law Enforcement Research Fund, Australian Government Cooperative Research Centre for Data to Decision, Lockheed Martin Australia, auDA Foundation, Government of South Australia, BAE Systems stratsec, Australasian Institute of Judicial Administration Incorporated, Australian Research Council, etc. 
\end{IEEEbiography}

\end{document}